\documentclass[letterpaper,twocolumn,10pt]{article}
\usepackage{usenix,epsfig, endnotes}

\usepackage{adjustbox}	
\usepackage{enumitem}
\usepackage{graphicx}
\usepackage{subcaption}
\usepackage{pifont}
\usepackage{booktabs}
\usepackage{makecell}
\usepackage{multirow}
\usepackage{amssymb}
\usepackage{array}    
\usepackage{colortbl}
\usepackage{tikz}
\usepackage{svg}
\usepackage{url}
\usepackage{soul}
\usepackage{threeparttable}
\usepackage{etoolbox}
\usepackage{tcolorbox}
\newtoggle{keepnotes}
\toggletrue{keepnotes}  
\usepackage[table,xcdraw]{xcolor}
\usepackage[table]{xcolor}
\usepackage{pgfmath}
\usepackage{arydshln}
\usepackage{hyperref}
\usepackage{tabularx}

\usepackage{xcolor}
\definecolor{niceblue}{HTML}{1f77b4}    
\definecolor{nicered}{HTML}{d26567}     
\definecolor{nicegreen}{HTML}{579997}   
\definecolor{niceorange}{HTML}{ff7f0e}  
\definecolor{nicepurple}{HTML}{9467bd}  
\definecolor{nicebrown}{HTML}{8c564b}   
\definecolor{nicegray}{HTML}{7f7f7f}    
\definecolor{myread}{HTML}{AF416A}
\definecolor{myorigin}{HTML}{FF874A}

\newcommand*{\eg}{\textit{e.g.,}}
\newcommand*{\ie}{\textit{i.e.,}}

\newcommand*{\emptycirc}[1][1ex]{\tikz\draw (0,0) circle (#1);}
\newcommand*{\halfcirc}[1][1ex]{%
    \begin{tikzpicture}
        \draw[fill] (0,0) -- (90:#1) arc (90:270:#1) -- cycle;
        \draw (0,0) circle (#1);
    \end{tikzpicture}%
}

\newcommand*{\fullcirc}[1][1ex]{\tikz\fill (0,0) circle (#1);}

\newcommand*{\emptydiamond}[1][1ex]{%
  \tikz\draw[scale=#1/6.5ex, line join=round] (0,1) -- (1,0) -- (0,-1) -- (-1,0) -- cycle;
}
\newcommand*{\halfdiamond}[1][1ex]{%
  \raisebox{-0.1ex}{
    \hbox{%
      \tikz[scale=#1/6.5ex, line join=round]{
        \path[fill] (0,1) -- (-1,0) -- (0,-1) -- cycle;
        \draw (0,1) -- (1,0) -- (0,-1) -- (-1,0) -- cycle;
      }%
    }%
  }
}

\newcommand*{\fulldiamond}[1][1ex]{%
  \tikz\fill[scale=#1/6.5ex] (0,1) -- (1,0) -- (0,-1) -- (-1,0) -- cycle;
}

\newcommand{\find}[1]{%
\begin{tcolorbox}[colback=black!5!white, boxrule=0.5pt, boxsep=2.1mm, left = 0pt, right=0pt, top=0pt, bottom=0pt]
\em #1
\end{tcolorbox}%
}

\newcommand{\update}[1]{\iftoggle{keepnotes}{\textcolor{black}{#1}}{}}
\newcommand{\ourbench}{\textsc{PatchEval}}

\begin{document}

\date{}

\title{\ourbench: A New Benchmark for Evaluating LLMs on Patching \\ Real-World Vulnerabilities}

\author{
\rm Zichao Wei$^{1\ddag *}$, Jun Zeng$^{2+}$, Ming Wen$^{1\ddag +}$, Zeliang Yu$^{1\ddag}$, Kai Cheng$^{1}$, Yiding Zhu$^{1}$,\\
\rm Jingyi Guo$^{1}$, Shiqi Zhou$^{2}$, Le Yin$^{2}$, Xiaodong Su$^{2}$, Zhechao Ma$^{2}$ \\[0.2em]
{$^1$Huazhong University of Science and Technology \quad $^2$ByteDance}
}

\renewcommand{\thefootnote}{\fnsymbol{footnote}}
\maketitle

\footnotetext[1]{Work done as an intern in ByteDance.{\footnotesize $^+$}Corresponding author.

\hspace{3pt}$^\ddag$National Engineering Research Center for Big Data Technology and System, Services Computing Technology and System Lab, 
Hubei Engineering Research Center on Big Data Security, Hubei Key Laboratory of Distributed System Security, School of Cyber Science and Engineering, HUST, China.


}

\thispagestyle{empty}

\subsection*{Abstract}
Software vulnerabilities are increasing at an alarming rate. 
However, manual patching is both time-consuming and resource-intensive, while existing automated vulnerability repair (AVR) techniques remain limited in effectiveness. 
Recent advances in large language models (LLMs) have opened a new paradigm for AVR, demonstrating remarkable progress.
To examine the capability of LLMs in AVR, several vulnerability benchmarks have been proposed recently. 
However, they still suffer from key limitations of outdated vulnerabilities, limited language coverage, unreliable patch validation, and insufficient reproducibility.

To overcome these challenges, we introduce \ourbench, a multilingual benchmark for Go, JavaScript, and Python, languages for which existing benchmarks remain unexplored. 
\ourbench{} curates a dataset of 1,000 vulnerabilities drawn from CVEs reported between 2015 and 2025, covering 65 distinct CWEs.
\update{A subset of 230 CVEs is further equipped with runtime sandbox environments, enabling patch verification through both security tests and functionality tests.}
To provide a systematic comparison of LLM-based vulnerability repair, we evaluate a series of state-of-the-art LLMs and agents, presenting an in-depth analysis that empirically yields key insights to guide future research in AVR. 

\section{Introduction}
The number of security vulnerabilities continues to increase at an unprecedented rate.
According to the National Vulnerability Database (NVD), 40,009 new CVEs were published in 2024 alone --- an increase of 38\% compared to 2023 and the highest annual count ever recorded~\cite{2024_cve}.
If left unpatched, these vulnerabilities can expose software systems or organizations to significant risks, including data leakage and financial loss~\cite{cve_loss}.
Patching vulnerabilities, however, remains a labor-intensive and time-consuming task~\cite{chen2022neural}.
\update{A typical workflow requires experts to
(1) localize the vulnerable code and identify its root cause,
(2) design a fix that is correct and minimally invasive,
and (3) validate the patch by evaluating whether the vulnerability is neutralized without introducing any regressions.}
Each step requires substantial manual effort.
As vulnerabilities accumulate faster than they can be patched, organizations face a growing backlog of unresolved issues.
Such high latency has long motivated research into automated vulnerability repair (AVR).

Early AVR approaches, ranging from template-based methods~\cite{huang2019using,rodler2021evmpatch} to search-based~\cite{zhang2022program,shi2022backporting}, constraint-driven~\cite{chen2017adaptive,chida2022repairing}, and more recently deep-learning-based techniques~\cite{chen2022neural,chi2022seqtrans}, achieve partial success but suffer from key limitations.
For example, template-based approaches lack generalization beyond predefined fixes.
Deep-learning–based approaches depend on large training datasets that are rarely available.
Recent advances in large language models (LLMs) open up a new paradigm. 
Trained on massive corpora of code, LLMs have demonstrated impressive performance in programming tasks~\cite{nam2024using}.
When equipped with external tools, agentic LLMs (or agents~\cite{DBLP:conf/nips/YangJWLYNP24,DBLP:conf/acl/ChenTDW0JPC025/localization_1}) are capable of addressing even more complex challenges like automated program repair (APR)~\cite{jimenez2023swe}.
This raises a natural question for the security community: Can LLMs generate valid patches for real-world vulnerabilities?

To address this question, researchers have recently begun to evaluate LLMs on the AVR task~\cite{DBLP:conf/msr/BuiSF22/vul4j,DBLP:conf/issta/WuJPLD0BS23/vjbench,DBLP:journals/corr/abs-2506-11791/secbench,DBLP:journals/corr/abs-2505-10494/CoV-Eval,DBLP:journals/corr/abs-2408-02153/arvo,DBLP:journals/corr/abs-2505-19395/vader,DBLP:journals/corr/abs-2506-11697/vul4c,DBLP:conf/naacl/WangLX25a/cvebench}, but existing evaluation benchmarks face several key limitations.
First, many benchmarks rely on outdated vulnerability datasets, containing CVEs that no longer reflect current attack vectors or software practices.
Second, they are narrow in programming language coverage, focusing primarily on C/C++ and Java, while overlooking other widely used languages such as Python, JavaScript, and Go, which dominate today’s web and cloud ecosystems and frequently appear in CVEs.
Third, patch validation in prior benchmarks often relies on human review --- subjective, inconsistent, and non-reproducible --- or on similarity metrics that compare LLM-generated patches with developer-authored fixes but cannot reliably verify whether the vulnerability has been removed.
More importantly, in the absence of manual ground truth labeling, developer-authored patches are often entangled with unrelated edits like functional refactoring~\cite{sun2025dispatch}, leading to misleading results in patch validation.
Finally, benchmark artifacts may lack reproducibility due to incomplete experimental environments or the absence of open-source availability.
These limitations leave several fundamental research questions unanswered:
How accurate are LLMs in generating patches for real-world vulnerabilities?
How do different prompt strategies affect the performance of LLMs?
Finally, can LLMs repair vulnerabilities that are hard to fix?

In this paper, we present \ourbench, a new benchmark specifically designed to evaluate LLMs in repairing real-world vulnerabilities.
\ourbench{} contains 1,000 vulnerabilities drawn from CVEs reported between 2015 and 2025, covering 65 CWE categories.
Unlike prior benchmarks that primarily focus on C/C++ and Java, \ourbench~emphasizes Python, Go, and JavaScript. 
These languages are selected because 
(1) they rank among the top ten most widely used languages in 2025\cite{top10_pl}, 
(2) they are also among the top ten languages containing the most number of CVEs~\cite{DBLP:conf/promise/BhandariNM21/cvefixes},
and (3) they have been largely overlooked in prior AVR benchmarks.
Each vulnerability in \ourbench~ is manually disentangled to separate security-relevant edits from unrelated functional or stylistic changes.
In addition, we provide runtime execution environments for 230 CVEs, each packaged in a dockerized sandbox to ensure reproducible patch validation and seamless agent interaction.
To reflect realistic AVR workflows~\cite{DBLP:conf/issta/WuJPLD0BS23/vjbench, nong2025appatchautomatedadaptiveprompting/appatch, PatchAgent, DBLP:journals/corr/abs-2501-03446/LLM4CVE}, \ourbench{} supports two evaluation scenarios:
(1) \textit{Patch Generation with Location Oracle}, where LLMs are given vulnerable function(s) and tasked solely with synthesizing fixes, and
(2) \textit{End-to-End Patch Generation}, where LLMs are required to localize and repair vulnerabilities, thereby simulating a complete AVR workflow in practice.

Patch validation is a long-standing challenge in evaluating AVR. 
To enable comprehensive evaluations of AVR, \ourbench~incorporates two complementary validation methodologies:
(1) \textit{static similarity matching}, which compares LLM-generated patches against real-world developer patches to assess consistency, and
(2) \textit{dynamic sandbox testing}, \update{which executes both security and functionality tests to verify that vulnerabilities are effectively neutralized without introducing regressions}.
Although static matching has been widely used in prior works~\cite{DBLP:journals/corr/abs-2501-03446/LLM4CVE, nong2025appatchautomatedadaptiveprompting/appatch, DBLP:conf/sigsoft/FuTLNP22/vulrepair, DBLP:conf/icse/ZhouKXH024/vulmaster}, it often yields misleading results --- syntactically similar patches may not correctly remove the vulnerability, whereas dissimilar ones can still be semantically correct. 
Following recent studies~\cite{wang2025can, nong2025appatchautomatedadaptiveprompting/appatch}, we also experiment by leveraging LLM-as-a-judge for patch validation. 
However, this approach proves unreliable as LLMs frequently overlook grammar errors and are often misled by superficial similarities~\cite{li2024llmsasjudgescomprehensivesurveyllmbased}.
\update{Consequently, dynamic sandbox testing remains our principal patch validation methodology, since it provides the most trustworthy evidence of correct vulnerability repair and regression-free functionality.}

We use \ourbench~to evaluate ten state-of-the-art LLMs, \update{one leading agents in APR, one code agent for general software development}, and one commercial agent in vulnerability repair.
Our experiments show that even the best-performing LLM or agent successfully repairs only \update{23.0\% (53/230)} of vulnerabilities --- far below the requirement for practical deployment.
More importantly, most successful repairs are concentrated in relatively simple patches, while existing models can hardly repair complex vulnerabilities that require substantial edits. 
Besides, our study also reveals that existing models exhibit strong complementarity in AVR: \update{98} vulnerabilities can be repaired by at least one model, yet the best-performing model can only repair \update{53}, and only \update{five} are fixed by all models.
\update{Our analysis also highlights the importance of accurate vulnerability localization and the need for agents tailored to AVR.}
In summary, we make the following contributions:


\begin{itemize}[leftmargin=12pt]
  \item We introduce \ourbench, a new benchmark for evaluating LLMs on real-world vulnerability repair tasks. 
  It contains 1,000 real-world vulnerabilities collected from CVEs reported between 2015 and 2025, spanning 65 distinct CWE categories. 
  Complementary to prior benchmarks that mainly focus on C/C++ and Java, \ourbench~emphasizes Python, Go, and JavaScript.

  \item We develop a fully automated and reproducible evaluation framework that integrates curated datasets with runtime execution environments. 
  Specifically, \ourbench~supports two practical evaluation scenarios: Patch Generation with Location Oracle and End-to-End Patch Generation.
  \update{To validate AVR-generated patches, \ourbench~enables both dynamic sandbox testing, which executes security and functionality tests, and static similarity matching, which compares candidate patches with developer-authored fixes.}

  \item We benchmark ten state-of-the-art LLMs, \update{one program repair agents, one general-purpose code agent}, and one commercial vulnerability repair agent.
  Our results reveal that even the best-performing approach achieves only a \update{23.0\%} success rate in patch generation.
  We also uncover the importance of vulnerability localization and patch validation in effective vulnerability repair.  
\end{itemize}

\section{Background and Motivation}

\subsection{Vulnerability Repair}
\label{section21}
\begin{table*}[t]
\centering
\begin{adjustbox}{width=1.0\textwidth}
\begin{threeparttable}
\scriptsize
\caption{Existing Benchmarks on Vulnerability Repair}
\label{related_bench}
\setlength\tabcolsep{7pt}     
\def\arraystretch{0.7}
\begin{tabular}{lcrrrrccccc}
\toprule
\multirow{2}{*}{\textbf{Benchmark}}  & 
\multirow{2}{*}{\makecell{\textbf{Published}\\\textbf{Year}}}& 
\multirow{2}{*}{\makecell{\textbf{Programming}\\\textbf{Language}}} & 
\multirow{2}{*}{\makecell{\textbf{Vulnerability}\\\textbf{Period}}} & 
\multirow{2}{*}{\textbf{\# CWE}} & 
\multirow{2}{*}{\textbf{\# Samples}} & 
\multicolumn{3}{c}{\textbf{Patch Validation}} & 
\multirow{2}{*}{\makecell{\textbf{Ground Truth}\\\textbf{Labeling}}} & 
\multirow{2}{*}{\makecell{\textbf{Artifact}\\\textbf{Reproduced}}} \\

\cmidrule{7-9}
& & & & & & \textbf{SM}\tnote{1} & \textbf{DT}\tnote{2} & \textbf{HR}\tnote{3} & & \\
\midrule
Big-Vul\cite{DBLP:conf/msr/FanL0N20/bigvul}    &2020  & C/C++ & 2009--2017 & 91 & 3,754                       & \ding{51}  & \emptydiamond[0.8ex]\tnote{4} &  \ding{55}       & \ding{55}                    & \fullcirc \\
ExtractFix\cite{DBLP:journals/tosem/GaoWDJXR21/extractfix} &2021  & C/C++  & 2012--2018 & 8 & 30                         & \ding{55}  &  \halfdiamond[0.8ex]\tnote{5} &  \ding{55}    & \ding{51}                   & \fullcirc \\
CVEfixes\cite{DBLP:conf/promise/BhandariNM21/cvefixes}   &2021  & 27 languages & 1999--2024 & 180  & 5,365               & \ding{51}  & \emptydiamond[0.8ex] &  \ding{55}         & \ding{55}                    & \fullcirc \\
PatchDB\cite{DBLP:conf/dsn/WangWF0J21/patchdb}    &2021 & C/C++ & 1999--2019 & 129 & 12,000                         & \ding{51}  & \emptydiamond[0.8ex] &  \ding{55}       & \ding{55}                    & \fullcirc \\
Vul4J\cite{DBLP:conf/msr/BuiSF22/vul4j}      &2022 & Java  & 2012--2021 & 25 & 79                           & \ding{55}  & \fulldiamond[0.8ex]\tnote{6} &  \ding{55}    & \ding{51}                   & \fullcirc \\
SmartFix\cite{DBLP:conf/sigsoft/SoO23/smartfix}   &2023 & Solidity  & 2018 & 3 & 361                            & \ding{55}  & \emptydiamond[0.8ex] &  \ding{51}     & \ding{51}                   & \halfcirc\tnote{7}   \\
VJBench\cite{DBLP:conf/issta/WuJPLD0BS23/vjbench}    &2023 & Java   & 2016--2022 & 37 & 42                          & \ding{51}  & \fulldiamond[0.8ex] &  \ding{55}     & \ding{51}                   & \fullcirc \\
Sec-Bench\cite{DBLP:journals/corr/abs-2506-11791/secbench}  &2025& C/C++ & 2016-2024 & 17 &200  & \ding{55}  & \halfdiamond[0.8ex] &  \ding{55} & \ding{51} & \fullcirc \\
CoV-Eval\cite{DBLP:journals/corr/abs-2505-10494/CoV-Eval}  &2025 & C/Python & NA\tnote{9} & 18 &477   & \ding{51}  & \emptydiamond[0.8ex] &  \ding{55}& \ding{51} & \fullcirc \\
ARVO\cite{DBLP:journals/corr/abs-2408-02153/arvo}       &2025 & C/C++ & 2016--2024 & NA\tnote{9} & 5,001                        & \ding{55}  & \halfdiamond[0.8ex] &  \ding{55}     & \ding{51}                   & \fullcirc \\
VADER\cite{DBLP:journals/corr/abs-2505-19395/vader}      &2025 & 15 languages& 2025  & 69 & 174                   & \ding{55}  & \emptydiamond[0.8ex] &  \ding{51}     & \ding{51}                   & \halfcirc   \\
Vul4C\cite{DBLP:journals/corr/abs-2506-11697/vul4c}      &2025 & C/C++ & 2010--2023 & 19 & 144                        & \ding{55}  & \fulldiamond[0.8ex] &  \ding{55}  & \ding{51}                   & \fullcirc \\
CVE-Bench\cite{DBLP:conf/naacl/WangLX25a/cvebench}  &2025 & Java/JS/Python/PHP & 2008--2022  & 103 & 509             & \ding{55}  & \halfdiamond[0.8ex] &  \ding{55}    & \ding{51}                   & \emptycirc\tnote{8}  \\
PatchAgent~\cite{PatchAgent} &2025 & C/C++ & 2012--2024 & NA\tnote{9} & 178                         & \ding{55}  & \fulldiamond[0.8ex] &  \ding{55}   & \ding{51}                   & \fullcirc \\ 
\midrule
\multirow{2}{*}{\textbf{\ourbench}} & \multirow{2}{*}{\textbf{2025}}& \multirow{2}{*}{Go/Python/JS} & \multirow{2}{*}{\textbf{2015 - 2025}} & \textbf{39} & \textbf{230}    & \ding{51}&\fulldiamond[0.8ex]&\ding{55} & \multirow{2}{*}{\ding{51}} & \multirow{2}{*}{\fullcirc} \\
&  & &  & \textbf{65} & \textbf{1,000} &\ding{51}&\emptydiamond[0.8ex]&\ding{55}  &     &   \\
\bottomrule
\multicolumn{11}{l}{\textsuperscript{1}~SM: Static Matching \textsuperscript{2}~DT: Dynamic Testing
\textsuperscript{3}~HR: Human Review
\textsuperscript{4}~\update{\emptydiamond No security or functionality test.}
\textsuperscript{5}~\update{\halfdiamond Only security test.}
\textsuperscript{6}~\update{\fulldiamond Both security and functionality test.}
}\\
\multicolumn{11}{l}{
\textsuperscript{7}~\halfcirc~Lack of automated reproduction support. \textsuperscript{8}~\emptycirc~Not open-sourced by the time of writing. 
\textsuperscript{9}~Synthetic vulnerabilities or without CWE assigned.
}\\
\end{tabular}
\end{threeparttable}  
\end{adjustbox}
\end{table*}
Vulnerability repair typically follows a three-stage workflow.

$\bullet~$\textit{Vulnerability Localization}:
    The first step is to determine where and why the vulnerability occurs. 
    This involves identifying the vulnerability type (e.g., SQL injection, Server-side request forgery), diagnosing its root cause, analyzing the execution context in which it is triggered, and pinpointing the exact code snippets that require patching.
        
$\bullet~$\textit{Patch Generation}:
    Once the vulnerable code has been identified, the next step is to construct a patch. 
    A high-quality patch should be minimal, changing only what is necessary, yet sufficiently to fully eliminate the security flaw.
    
$\bullet~$\textit{Patch Validation}:
    The final stage is to verify that the patch truly resolves the vulnerability without introducing new issues like syntactic errors.
    Common validation strategies fall into three categories. 
    \textit{Human Review}:
    security experts manually inspect the patch and its documentation to confirm that the intended security invariant has been restored.
    \textit{Static Matching}: 
    when a ground-truth fix is available (e.g., from a developer-authored commit), the candidate patch can be compared against it to verify equivalence.
    \textit{Dynamic Testing}: 
    \update{the patched program is executed against Proof-of-Concept exploits (i.e., security test) and unit tests (i.e., functionality test) to verify that the vulnerability can no longer be triggered and that the original functionality remains intact.}

As vulnerabilities continue to proliferate at an unprecedented pace,  traditional human-driven vulnerability repair has become  unsustainable. 
To address this challenge, automated vulnerability repair (AVR) has attracted huge attention from the community, a paradigm that seeks to automate the entire vulnerability repair life cycle with minimal human effort~\cite{nong2025appatchautomatedadaptiveprompting/appatch, DBLP:conf/sigsoft/FuTLNP22/vulrepair, DBLP:conf/issta/WuJPLD0BS23/vjbench, DBLP:conf/sigsoft/SoO23/smartfix, DBLP:conf/sp/PearceTAKD23/exam_zero_shot_sp, DBLP:journals/corr/abs-2405-04994/NAVRepair}.
Two task formulations are widely adopted in AVR, which differ in whether the location of the vulnerability is assumed to be known~\cite{DBLP:journals/corr/abs-2401-16185/LLM4Vuln, DBLP:conf/sigsoft/SoO23/smartfix,DBLP:conf/sp/PearceTAKD23/exam_zero_shot_sp}:
\textit{Patch Generation with Location Oracle}: 
In this setting, the vulnerable code region (e.g., functions) is provided.
The AVR system is only tasked to generate a patch and correctly integrate it into the original software.
\textit{End-to-End Patch Generation}:
This setting requires the AVR system not only to generate a patch but also to first localize the vulnerable code, which yields a more realistic evaluation for practical deployment. 

\subsection{Patch Generation with LLMs}

Large language models (LLMs) now stand at the forefront of code intelligence~\cite{nam2024using, DBLP:conf/icse/NamMHVM24, DBLP:journals/corr/abs-2308-01240/evaluting_code_comprehension_and_generation}.
Trained on massive corpora of source code drawn from a wide range of software systems, they go far beyond syntactic understanding. 
The strengths of LLMs make them a natural fit for complex programming tasks like Automated Program Repair (APR)~\cite{nong2025appatchautomatedadaptiveprompting/appatch,PatchAgent,DBLP:journals/corr/abs-2501-03446/LLM4CVE,DBLP:journals/corr/abs-2401-16185/LLM4Vuln,DBLP:journals/corr/abs-2405-04994/NAVRepair,DBLP:conf/issta/PanWL00L24/ppathf,10.1145/3691620.3695349/contract_thinker, DBLP:conf/sp/PearceTAKD23/exam_zero_shot_sp, DBLP:conf/sigsoft/SoO23/smartfix}, which seeks to automatically fix software faults ranging from functional bugs to compilation errors. 
Unlike general bugs, vulnerabilities must be repaired in a way that not only preserves functional correctness but eliminates exploitable behavior under an explicit threat model. 
This additional security constraint makes AVR inherently more complex and higher-stakes than traditional APR, since an inadequate repair can leave software systems exposed to real-world attacks.
Recent work has already demonstrated the potential of LLMs for AVR.
For instance, Pearce et al.~\cite{DBLP:conf/sp/PearceTAKD23/exam_zero_shot_sp} first examine zero-shot vulnerability repair using commercial, open-source, and locally fine-trained LLMs, finding near-perfect success on synthetic vulnerabilities but significantly weaker performance on real-world cases.
SmartFix~\cite{DBLP:conf/sigsoft/SoO23/smartfix} improves patch quality by adopting a generate-and-verify strategy that iteratively proposes candidate patches and validates them with a safety verifier.
APPatch~\cite{nong2025appatchautomatedadaptiveprompting/appatch} introduces an adaptive prompting technique that guides LLMs to reason about the root causes of vulnerabilities.
Agent-based approaches further extend this direction:
PatchAgent~\cite{PatchAgent} integrates external tools such as language servers and patch verifiers to mimic human-like reasoning during vulnerability repair, achieving strong results on C/C++ vulnerabilities.


\subsection{Benchmarking Vulnerability Repair}
\label{section23}
To enable fair evaluation of LLMs on AVR, recent studies have created benchmarks that contain real-world vulnerabilities with standardized evaluation protocols, as summarized in Table~\ref{related_bench}. 
For instance, Extractfix~\cite{DBLP:journals/tosem/GaoWDJXR21/extractfix} creates a dataset of 30 manually verified C/C++ vulnerability instances collected from 2012 to 2018, along with a security test suite for verifying LLM-generated patches.
VADER~\cite{DBLP:journals/corr/abs-2505-19395/vader} contains 174 manually verified vulnerability instances spanning 15 languages; candidate patches are scored on a 0–10 scale to quantify repair quality. 
Unfortunately, existing benchmarks suffer from several key limitations, as discussed below:

\textit{Outdated Vulnerabilities}:
Most vulnerabilities used in the benchmarks are constructed from CVEs reported years ago and therefore fail to capture more recent attack vectors.
For instance, ExtractFix~\cite{DBLP:journals/tosem/GaoWDJXR21/extractfix} and SmartFix~\cite{DBLP:conf/sigsoft/SoO23/smartfix} only include vulnerabilities up to 2018, while VJBench~\cite{DBLP:conf/issta/WuJPLD0BS23/vjbench} and CVE-Bench~\cite{DBLP:conf/naacl/WangLX25a/cvebench} contain vulnerabilities reported before 2022. 
Although VADER~\cite{DBLP:journals/corr/abs-2505-19395/vader} contains vulnerabilities up to 2025, it lacks dynamic validation support as well as automated reproduction support.
As such, these benchmarks cannot reliably reveal whether LLMs are capable of repairing vulnerabilities that reflect today’s rapidly evolving threat landscape.

\textit{Limited Coverage of Programming Languages}:
Existing datasets remain heavily centered on C/C++ and Java.
Even the most up-to-date benchmarks, such as SEC-Bench~\cite{DBLP:journals/corr/abs-2506-11791/secbench}, Vul4C~\cite{DBLP:journals/corr/abs-2506-11697/vul4c}, and PatchAgent~\cite{PatchAgent}, still focus on C/C++, while widely used datasets like Vul4J~\cite{DBLP:conf/msr/BuiSF22/vul4j} and VJBench~\cite{DBLP:conf/issta/WuJPLD0BS23/vjbench} collect only Java vulnerabilities.
However, the software ecosystem has increasingly shifted toward languages like Python, JavaScript, and Go, which dominate today’s software landscape, like AI, web development, and cloud-native infrastructure.
These languages differ not only in syntax but also in the types of vulnerabilities.
For example, C/C++ benchmarks largely capture memory safety issues (e.g., \textit{buffer overflows} and \textit{use-after-free}), but Python/JavaScript/Go often suffer from \textit{authentication flaws}, \textit{cross-site scripting}, \textit{cloud service mis-configurations} and so on.
Accordingly, the continued reliance on benchmarks centered on C/C++/Java significantly constrains the generalizability of evaluation results.

\textit{Unreliable Patch Validation}:
\update{The gold standard for patch validation is dynamic execution~\cite{DBLP:journals/tosem/GaoWDJXR21/extractfix, PatchAgent, DBLP:journals/corr/abs-2506-11697/vul4c, DBLP:conf/issta/WuJPLD0BS23/vjbench}: rerunning both Proof-of-Concepts and unit test in a reproducible environment to verify whether the vulnerability can still be triggered without regressions.}
Nonetheless, many studies bypass dynamic validation, as constructing verification oracles and execution environments is both time-intensive and resource-demanding.
To compensate, existing benchmarks fall back on alternative methods,  most commonly human review~\cite{DBLP:journals/corr/abs-2505-19395/vader, DBLP:conf/sigsoft/SoO23/smartfix}, static similarity matching~\cite{DBLP:journals/corr/abs-2501-03446/LLM4CVE, DBLP:conf/sigsoft/FuTLNP22/vulrepair}, and LLM-based judges~\cite{nong2025appatchautomatedadaptiveprompting/appatch}. 
While these approaches offer greater scalability and lower cost, each comes with well-known flaws.
For example, human reviews are highly subjective and often inconsistent across evaluators.
Static similarity testing, such as Exact Match and CodeBLEU~\cite{DBLP:journals/corr/abs-2405-04994/NAVRepair, DBLP:journals/corr/abs-2501-03446/LLM4CVE}, largely conflate syntactic resemblance with semantic correctness, so a patch may appear similar to the ground-truth fix yet still leave the vulnerability exploitable. 
As LLM-based judges do not execute code, their assessment is bounded by model reasoning, making them prone to both false positives and negatives. 
A case study is provided in Appendix as shown in Figure~\ref{llm_judge_case} to illustrate this issue. 

\textit{Lack of Reproducibility}:
Reproducibility in AVR benchmarks depends on two properties: (1) the degree of automation in vulnerability repair and (2) the accessibility of code and data.
When measured against the USENIX artifact open-source criteria~\cite{sec_badges}, many benchmarks yet fall short.
For example, VADER~\cite{DBLP:journals/corr/abs-2505-19395/vader} contributes a multi-language vulnerability repair dataset but relies on manual patch verification, which limits automation and undermines reproducibility.
In contrast, CVE-Bench~\cite{DBLP:conf/naacl/WangLX25a/cvebench} provides a fully automated repair validation pipeline, but its lack of open-source limits accessibility and prevents the community from evaluating new approaches.

To overcome these limitations, we introduce \ourbench, a new benchmark designed to evaluate LLMs on repairing real-world vulnerabilities with the following properties:
\ding{202} Up-to-Date Vulnerabilities: 
The dataset is constructed from recent real-world vulnerabilities to capture the latest attack vectors. 
This enables \ourbench~to evaluate whether LLMs can handle vulnerabilities that are not present in earlier datasets and better reflect the current threat landscape.
\ding{203} Diverse Language Coverage: 
\ourbench~extends beyond the traditional focus on C/C++/Java by incorporating programming languages that dominate modern software systems but remain underrepresented in existing benchmarks. 
\ding{204} Valid Patch Validation: 
We equip vulnerabilities with both similarity-based and execution-based patch validation. 
This dual approach ensures that LLM-generated patches are assessed not only for syntactic correctness but also for security effectiveness.
\ding{205} Repository-Level Evaluation: 
\ourbench~operates at the repository level to support end-to-end vulnerability repair. 
This setup allows an LLM to interact with the full codebase --- navigating the project, localizing vulnerable code, and integrating patches into the context --- closely mirroring how developers repair vulnerabilities.
\ding{206} Reproduced Artifact:
All datasets, along with implementations for patch generation and validation, are publicly released. 
The evaluation framework is designed to be user-friendly, making it straightforward to rerun evaluations or integrate new approaches into the benchmark. 
This openness not only ensures that results can be independently reproduced but also promotes transparency and facilitates future research for extension.

\begin{figure*}[t!]
    \centering
    \includegraphics[width=0.9\textwidth]{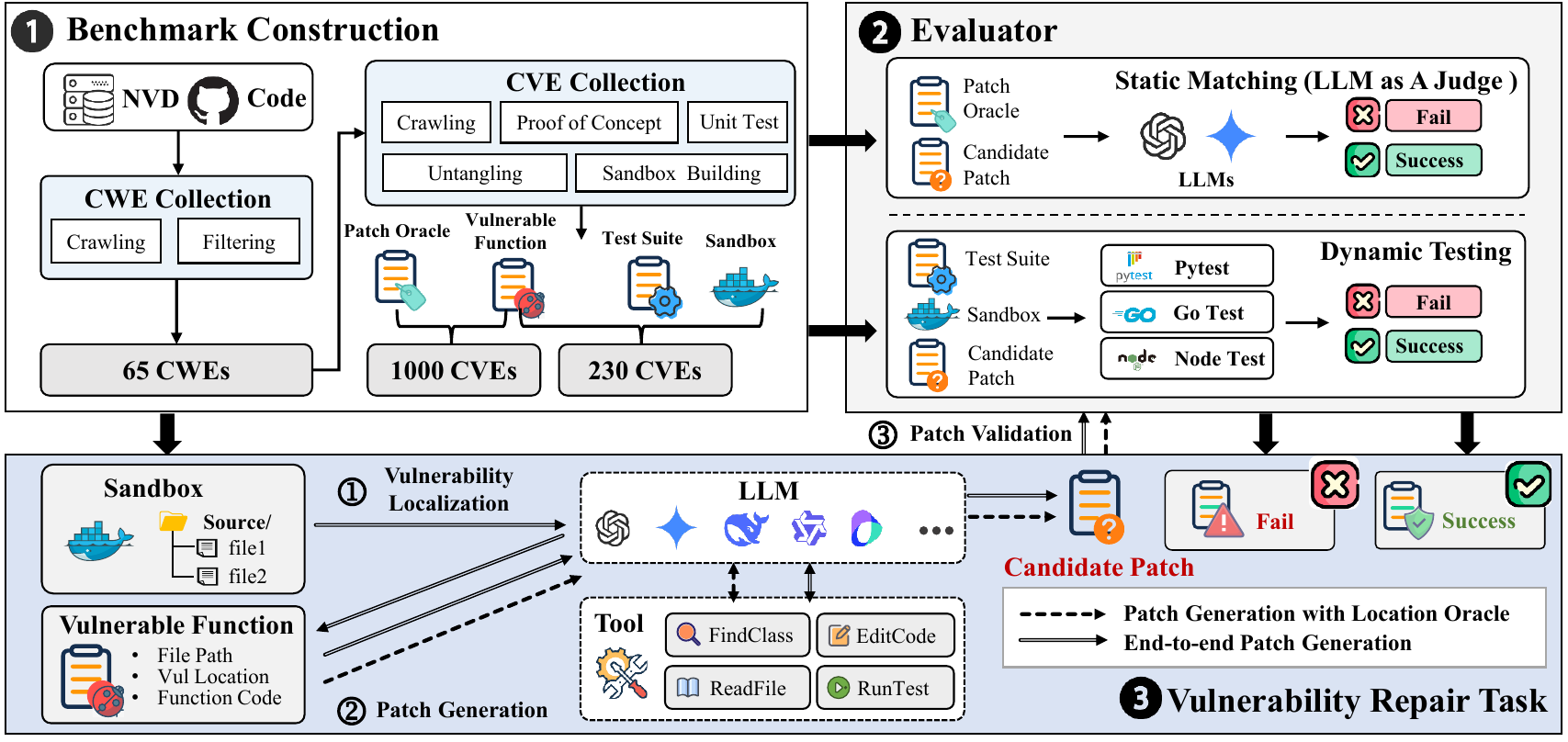}
    \caption{Overview of \ourbench }
    \label{overview}
\end{figure*}
\section{\ourbench{}}


Figure~\ref{overview} presents an overview of \ourbench, a vulnerability repair evaluation framework designed to work with any LLM and code agent.
\ourbench~consists of three primary phases:
Benchmark Construction (Section~\ref{benchmark construction}):
includes 1,000 real-world vulnerabilities reported as CVEs between 2015 and 2025.
Evaluator (Section~\ref{Evaluator}):
incorporates two complementary patch validation methodologies: \textit{static similarity matching} and \textit{dynamic sandbox testing}.
Task Formulations (Section~\ref{task_formulation}): supports 
 two standard formulations of vulnerability repair: \textit{patch generation with the location oracle} and \textit{end-to-end patch generation}. 

\begin{figure}[t!]
    \includegraphics[width=0.46\textwidth]{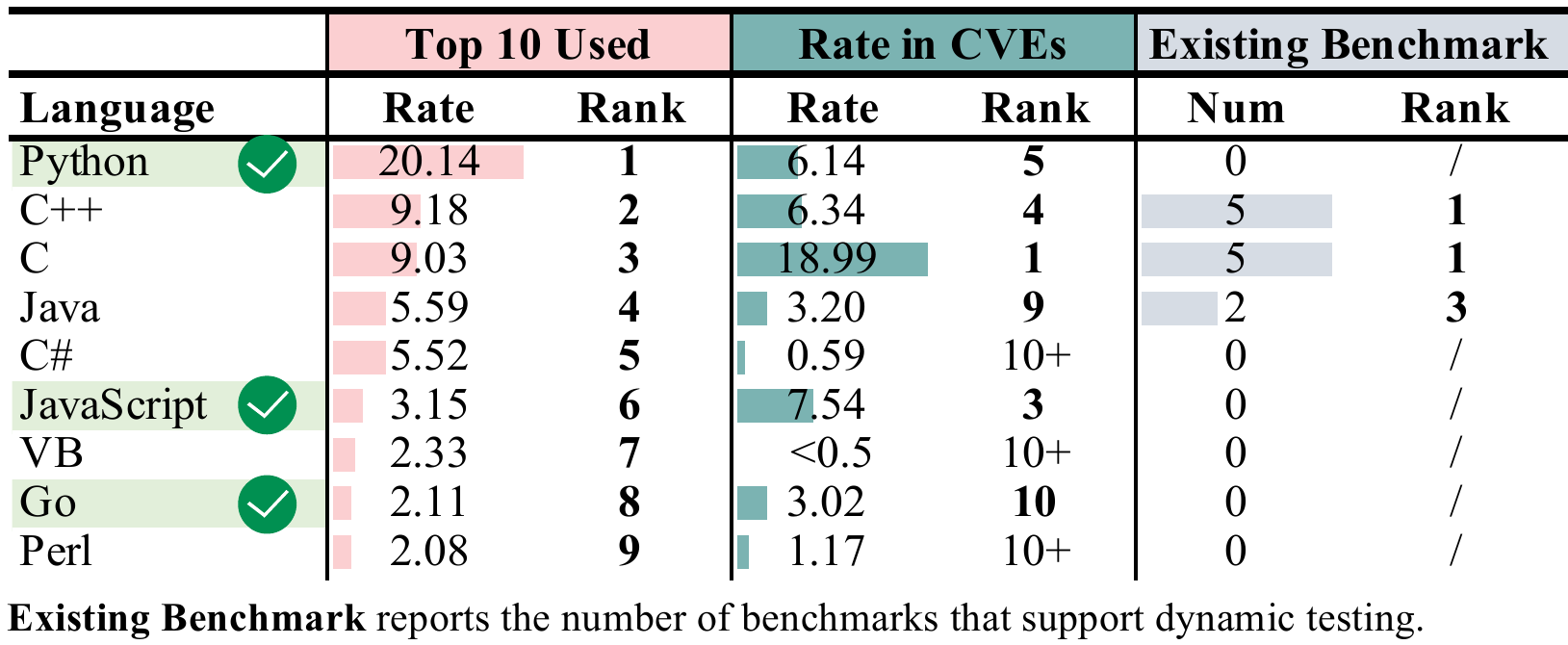}
    \caption{Programming Language Selection Criterial}
    \label{language selection}

\end{figure}

\subsection{Benchmark Construction} \label{benchmark construction}
Before constructing a representative benchmark for AVR, we first identify the target programming languages based on three criteria: 
ranking among the top ten most popular languages;
ranking among the top ten languages with the highest number of documented CVEs; and 
being largely overlooked in prior AVR benchmarks, where the lack of coverage prevents systematic evaluation.
Figure~\ref{language selection} summarizes the statistics underlying these criteria.
The first column (Top 10 Used) lists the ten most widely adopted programming languages as of August 2025~\cite{top10_pl}. 
The second column (Rate in CVEs) ranks languages by their share of reported CVEs up to July 2024~\cite{DBLP:conf/promise/BhandariNM21/cvefixes}.
We exclude non-programming languages such as Markdown and HTML. 
The third column shows the number of existing benchmarks in Table~\ref{related_bench} that support dynamic testing.

Our selection begins with languages that appear in both the top ten most widely used and the top ten with the highest number of documented CVEs. 
This intersection yields six candidates: Python, C, C++, Java, JavaScript, and Go.
To further address gaps in prior work, we then focus on those languages that largely remain unexplored, which finally led us to Python, JavaScript, and Go.
This choice broadens the scope of AVR evaluation beyond the traditional focus on C/C++ and Java. 
In particular, while C/C++ and Java remain central to systems programming, Python, JavaScript, and Go dominate modern computing ecosystems --- Python as the backbone of AI, JavaScript as the foundation of the web, and Go as the language of choice for cloud-native infrastructure.
Evaluating AVR in these languages thus provides a more representative view of the challenges in contemporary vulnerability repair.

\textbf{CWE Collection.}
To select the scope of vulnerability types for \ourbench, we start from the full set of 943 Common Weakness Enumeration (CWE) entries listed on the official CWE website as of August 2025.\footnote{https://cwe.mitre.org/data/definitions/1000.html} 
From this pool, we next apply a filtering pipeline to retain only CWEs that are both practically significant and experimentally viable.
Our first step is to prioritize prevalence: we select only CWEs linked to more than 100 CVEs, narrowing the pool to 110 high-impact candidates. 
This set is further refined using two suitability criteria as follows.
Language relevance: the vulnerability type must be applicable across all three target languages (Python, JavaScript, and Go). 
Data sufficiency: each CWE must be linked to at least ten CVEs to provide adequate samples for vulnerability repair. 
A CVE is considered valid only if its developer-provided patch is publicly accessible.
Applying these filters yields a final set of 65 CWEs, which defines the scope of~\ourbench.

\textbf{CVE Collection}
From these 65 CWEs, we collect 1,000 CVEs from the National Vulnerability Database (NVD)\footnote{https://nvd.nist.gov} to construct our vulnerability dataset.
The accuracy of vulnerability information, such as the vulnerable code and its corresponding patch, forms the foundation of a high-quality dataset for AVR.
Publicly disclosed CVEs, together with their officially linked fix commits in NVD, provide a solid starting point.
However, a recurring challenge is the prevalence of tangled commits --- commits that bundle the security fix with unrelated modifications such as refactoring, functionality development, and documentation updates~\cite{DBLP:conf/icse/WangHGWC024/reposvul,DBLP:conf/icse/Luo0024,DBLP:conf/sigsoft/PartachiDAB20}.
If left unfiltered, these tangled commits introduce significant noise.
This contamination not only skews evaluation results but also limits the generalization of vulnerability repair methods in real-world settings.
To build a trustworthy ground truth of vulnerability patches, we follow a minimal patch principle~\cite{sun2025dispatch}: each patch must include only the code strictly necessary to eliminate the vulnerability. 
To do so, a team of four security researchers, each with more than three years of experience in vulnerability analysis, carries out a three-step manual untangling process:
(1) inspecting each candidate fix’s \textit{git diff} to identify the security-relevant changes,
(2) cross-referencing the intended fix with the corresponding CVE description and CWE entry,
and (3) extracting only the lines essential for eliminating the vulnerability, while discarding unrelated edits such as code refactoring and functionality developments.
Through this manual labeling process, we obtain 1,000 minimized security patches, each preserving only the code changes necessary to eliminate a vulnerability.

To ensure dataset quality, a second-pass review of 1,000 CVEs is conducted by an independent team of eight senior security experts, resulting in 103 patches to be revised due to untangling errors or inappropriate fixes (e.g., deprecating APIs). 
This review not only enhances the reliability of our dataset but also finds inaccuracies in NVD records~\cite{cve_2022_0406}, revealing quality issues in widely used vulnerability repositories.



Starting with a dataset of 1,000 CVEs, our next objective is to construct Proof-of-Concepts (PoCs) to enable dynamic security verification.
We begin by analyzing the test cases committed alongside vulnerability patches.
Prior work finds that test cases --- crafted to expose a vulnerability --- often appear in the same or a nearby Git commit as the corresponding fix~\cite{DBLP:conf/issta/WuJPLD0BS23/vjbench,DBLP:conf/naacl/WangLX25a/cvebench}. 
Guided by this observation, we conduct a manual review of all 1,000 CVEs to identify new tests that explicitly exercise the security fix. 
For each candidate patch, we rebuild both the vulnerable (pre-patch) and the fixed (post-patch) versions of the software and execute the test suite. 
A test qualifies as a valid PoC if it fails on the pre-patch version but succeeds on the post-patch version.
More specifically, each test is executed inside an isolated sandbox environment.
We implement these sandboxes using Docker containers, which bring two key benefits:
(1) reproducibility: by running security tests in the same container image, containers guarantee consistent runtime behavior across executions, 
and (2) scalability: unlike virtual machines, containers can be created and destroyed efficiently, minimizing overhead when orchestrating hundreds of sandbox environments in parallel.

\update{In practice, many security patches either lack test cases or include those that do not trigger the vulnerability. 
Among 1,000 CVEs, we identify public PoCs for 442 and validate them by confirming that each PoC fails before patching but succeeds afterward. 
Allocating two hours per PoC for vulnerability reproduction, we obtain 192 PoCs, as many attempts fail due to incomplete environment specifications.
For the remaining 558 CVEs without existing PoCs,  we manually construct PoCs from scratch.
The process begins with gathering all public information on CVEs, such as official NVD entries and relevant technical reports. 
We then identify the root cause of each vulnerability, analyze its triggering conditions, and implement an exploit that reliably reproduces the issue.
Given the variable quality of available CVE information, PoC reconstruction can be straightforward or require substantial debugging.
We therefore impose a two-hour budget on PoC reconstruction per CVE and successfully create 38 PoCs. 
Following the two steps described above, we finally construct PoC tests for 230 vulnerabilities.}

\update{To further collect unit tests for these CVEs, we manually inspect the corresponding GitHub repository to identify test cases that exercise the functionality affected by the patch.
Specifically, we prioritize test cases that are co-located with or depend on the patched source files, expanding the search outward as necessary.
This locality-driven strategy effectively narrows the search space while maintaining high coverage of unit tests suitable for regression validation.}

\textbf{Final Datasets}. 
Eventually, we curate two complementary datasets to evaluate the performance of LLMs on AVR.
\textit{Equivalence Validation Dataset}:
This dataset contains 1,000 CVEs disclosed between 2015 and 2025, covering 65 CWEs across three major programming languages (i.e., Python, JavaScript, and Go). 
Each sample is manually labeled through careful disentangling of vulnerability-fixing commits.
The dataset is intended to evaluate whether an AVR-generated patch is equivalent in security intention to the official developer-provided fix.
\textit{PoC Validation Dataset}:
To enable more reliable evaluation of vulnerability repair, we construct a second dataset that offers high-fidelity dynamic patch validation.
This dataset consists of 230 vulnerabilities, each paired with a sandbox environment and manually crafted test functions. 

\subsection{Evaluator} \label{Evaluator}
Patch validation is typically conducted using three strategies.
\textit{Static Matching} compares a candidate patch against the developer-provided ground truth to evaluate both syntactic and semantic equivalence.
\textit{Dynamic Execution} runs the patched program against PoCs and monitors its runtime behavior.
A patch is considered correct if the vulnerability can no longer be triggered by the PoCs.
Manual Review relies on the expertise of security analysts to judge patch correctness. 
While useful, this strategy is inherently subjective and difficult to scale.
Given the trade-offs, \ourbench~focuses on static matching and dynamic execution, as these methods enable reproducible and fully automated evaluation.

\textbf{Static Matching.}
Static matching evaluates each candidate patch by measuring its similarity to the ground-truth patch.
In practice, three approaches are often used to measure patch similarities.
The first is \textit{exact match}~\cite{DBLP:conf/sigsoft/FuTLNP22/vulrepair}, which considers a patch correct only if it is identical to the ground truth. 
The second is \textit{similarity quantification}~\cite{DBLP:journals/corr/abs-2501-03446/LLM4CVE, DBLP:journals/corr/abs-2405-04994/NAVRepair}, which computes a similarity score~CodeBLEU~\cite{DBLP:journals/corr/abs-2501-03446/LLM4CVE} to estimate whether a candidate fix aligns with the ground truth.
Both methods, however, operate at the syntactic level and may incorrectly reject patches that differ in syntax but are semantically equivalent.

Recent studies show that LLMs can go beyond syntax-level comparison and capture the underlying semantic intent of code.
As a result, they demonstrate near-human performance in evaluating software engineering tasks like code translation and generation~\cite{wang2025can}.
Similarly, LLMs have also been adopted as a judge in patch validation tasks~\cite{nong2025appatchautomatedadaptiveprompting/appatch}.
In this setting, both the candidate patch and the ground truth are provided to the LLM via carefully designed prompts that also include vulnerability context. 
LLM is then instructed to reason about whether the candidate constitutes a valid fix and to output both a binary verdict (\ie~\textit{success/fail}) and an explanatory rationale supporting its decision.
That said, LLM as a judge remains limited by its textual perspective.
For example, LLM usually cannot detect runtime issues such as grammar errors.

\textbf{Dynamic Testing.}
Dynamic testing effectively addresses the limitations of static matching by directly verifying whether a vulnerability can still be exploited after a patch is applied. 
This approach requires both an isolated, executable environment and a corresponding security test, typically in the form of a PoC. 
A valid PoC produces different outcomes on the pre- and post-patch versions of the code: it triggers the vulnerability in the pre-patch version but is neutralized in the post-patch version. 
Because these tests are specifically designed by developers to confirm that a vulnerability has been fixed, their execution results provide a reliable indicator of patch correctness.

In \ourbench, dynamic evaluation begins by initializing a sandboxed environment that uses the vulnerable pre-patch version as the baseline. 
The security test is first executed to confirm the existence of the vulnerability. 
Afterward, the candidate patch generated by an LLM is applied, and the same test is rerun to determine whether the vulnerability has been successfully eliminated. 
\update{To further ensure functional correctness, a complementary suite of unit tests is subsequently executed to detect any regressions introduced by the patch.}
All dynamic tests --- including both security and unit tests --- are executed inside isolated sandbox environments to ensure security, reproducibility, and reliability, while also preventing potential contamination on the host system.


\subsection{Task Formulation}

\label{task_formulation}
\subsubsection{Patch Generation with Location Oracle} \label{patch_generation_with_location_oracle}
Our initial objective is to evaluate the AVR capabilities of existing LLMs with known vulnerability locations. Following the existing researches~\cite{nong2025appatchautomatedadaptiveprompting/appatch, DBLP:conf/issta/WuJPLD0BS23/vjbench, DBLP:journals/corr/abs-2405-04994/NAVRepair}, our experimental setup involves directly providing the vulnerable function to LLMs as a location oracle.
This approach isolates the repair task from the localization task, in which case the LLM does not need to search for the vulnerability within an entire codebase. Such a setting allows us to focus our evaluation squarely on the code-fixing capabilities of LLMs. 

\begin{figure}[t!]
    \centering
    \includegraphics[width=0.45\textwidth]{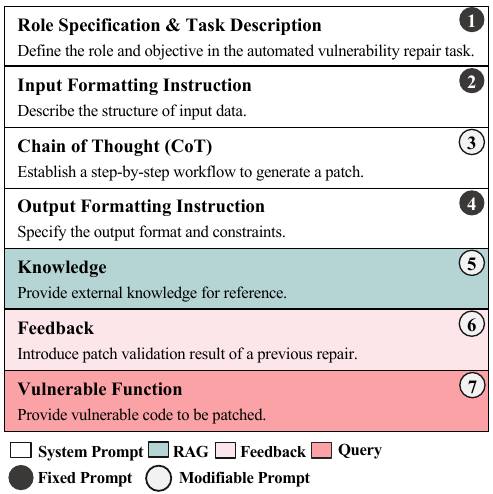}
    \caption{Prompt for Vulnerability Repair }
    \label{prompt}
\end{figure}

\textbf{Prompt Strategy Design.}
To explore how different prompt strategies affect the performance of LLMs, \ourbench{}~designs and evaluates several different
prompt configurations, as shown in Figure~\ref{prompt}. 
Specifically, each prompt is constructed from the following components: 
\ding{202} Role and Task: Sets the persona of the model (e.g., ``\textit{You are a code security expert}'') and defines the primary objective of the task.
\ding{203} Input Formatting Instruction: Describes the structure of the input data, instructing the model to understand the purpose of each field (e.g., CVE  description, one-shot example).
\ding{204} Chain of Thought (CoT): Instructs the model to follow a specific reasoning process, such as first analyzing the persona of the vulnerability, then proposing a repair strategy, and finally generating the patched code.
\ding{205} Output Formatting Instruction: Specifies the desired format for the output (e.g., a JSON object) and any constraints the generated patch must meet.
\ding{206} Knowledge: Provides supplementary information to improve repair accuracy. 
This includes the CWE/CVE details of the target vulnerability, along with a historical example of a similar vulnerability (under the same CWE and language) whose CVE predates the one under evaluation to prevent data leakage.
\ding{207} Feedback: For iterative repairs, this information is updated based on the outcome of patch validation (e.g., compiler errors or failing PoC cases).
\ding{208} Vulnerable Function: Supplies the target code region for repair—typically the vulnerable function, with function signature, and function body as context to enable correct patch generation.

Based on the components above, we create four experimental settings as shown in Table~\ref{experiment_setting}: 
(1) \textit{Full Prompt (S1.1)}: the baseline configuration, which includes all components but feedback for a single-turn repair, providing both CoT guidance and domain knowledge. 
(2) \textit{Without CoT~(S1.2)}: removes explicit CoT instructions to evaluate the performance of LLM without reasoning guidance. 
(3) \textit{Without Knowledge~(S1.3)}: skips the domain knowledge to measure its impact on the quality of the repair. 
(4) \textit{With Feedback~(S1.4)}: enables a multi-turn setting where LLM can leverage patch verification results as feedback to refine its solution.

\begin{table}[t]
\centering
\begin{adjustbox}{width=0.45\textwidth}
\begin{threeparttable}          
\caption{Experiment Settings on \ourbench}
\label{experiment_setting}
\footnotesize
\def\arraystretch{0.85}
\begin{tabular}{ p{2.6cm}cccc }
\toprule
 \textbf{Settings} & \multicolumn{1}{m{0.9cm}}{\textbf{Location}} & \multicolumn{1}{p{1.2cm}}{\textbf{Knowledge}} & \multicolumn{1}{p{0.4cm}}{\textbf{CoT}} & \multicolumn{1}{p{1.2cm}}{\textbf{Feedback}}  \\
\midrule
S1.1 {\textbf{W/ Location}$^*$} & \ding{51} & \ding{51} & \ding{51} & \ding{55}   \\
 S1.2 {w/o CoT} & \ding{51} & \ding{51} & \ding{55} & \ding{55}   \\
 S1.3 {w/o Knowledge} & \ding{51} & \ding{55} & \ding{51} & \ding{55}  \\
 S1.4 {w/ Feedback} & \ding{51} & \ding{51} & \ding{51} & \ding{51}   \\
\midrule
 S2.1 {\textbf{W/o Location}$^*$}  & \ding{55} & \ding{51} & \ding{51} & \ding{55}   \\
  S2.2 {w/ Feedback}  & \ding{55} & \ding{51} & \ding{51} & \ding{51}   \\
\bottomrule
\multicolumn{5}{l}{S1.x denotes patch generation with location oracle;} \\
\multicolumn{5}{l}{S2.x denotes end-to-end patch generation. $^*$ is the default setting.}
\end{tabular}

\end{threeparttable}       

\end{adjustbox}
\end{table}


\textbf{Vulnerability Localization Impact.}
To better approximate real-world scenarios where the exact location of a vulnerability is often unknown, we highlight the role of fault localization and present the first large-scale empirical study on its impact in LLM-based AVR to our knowledge. 
Although prior work~\cite{DBLP:conf/acl/ChenTDW0JPC025/localization_1, DBLP:conf/aaai/OrvalhoJM25/localization_2, DBLP:conf/issta/WuJPLD0BS23/vjbench} has extensively explored vulnerability localization, accurately pinpointing faulty lines remains challenging. 
Motivated by its difficulty and importance for AVR, we investigate how the precision of provided locations influences repair performance. 
Specifically, 
\ourbench{}~designs four sets of localization results with various precisions, all within the context of the provided vulnerable function:
(1) \textit{Without Line}: LLM receives the entire vulnerable function without any hint of vulnerable lines. 
(2) \textit{Precise Line}: LLM is given the exact lines where the official patch is applied.
(3) \textit{Approximate Line}: LLM is directed to a random line within five lines of the true vulnerable location. 
(4) \textit{Imprecise Line}: LLM is directed to a random line more than five lines away from the true vulnerable location.
These experiments simulate vulnerability localization at different levels of precision to assess its impact on the effectiveness of AVR.

\subsubsection{End-to-End Patch Generation} 
\label{end to end patch generation}
Recent agent-based approaches~\cite{PatchAgent,nong2025appatchautomatedadaptiveprompting/appatch,DBLP:conf/nips/YangJWLYNP24} represent a major advance, extending beyond isolated functions to operate on entire code repositories.
These approaches manage to complete the entire repair workflow, from localizing a vulnerability to generating a patch, in a process that closely mirrors real-world practices.
The workflow takes two inputs: the vulnerable code repository and a description of the vulnerability. 
The agent-based LLM then interacts with a sandbox environment using a suite of built-in tools. 
For example, it may first query the repository to identify relevant files and then inspect their contents to locate the vulnerable function. 
Once the vulnerability is identified, the agent applies a predefined editing tool to modify the code and apply a fix. 
It may further run tests to validate the modifications before finalizing the patch.

The key difference between this End-to-End setting and the setting in Section~\ref{patch_generation_with_location_oracle} is that the vulnerability location is not provided in advance.
Instead, the agent must proactively identify it using the available tools. 
This design grants the LLM greater flexibility and more closely simulates the workflow of a human developer, who investigates the surrounding context of the vulnerable code before constructing a correct patch.
Accordingly, \ourbench~designs the following experimental settings:
(1) \textit{With Location (S1.1)}: the agent is provided with the vulnerable functions directly;
(2) \textit{End-to-End (S2.1)}: the agent is given only the vulnerable repository and must autonomously perform both localization and repair;
(3) \textit{With Feedback (S2.2)}: the agent additionally receives feedback from the patch verification to guide iterative repair.

\section{Evaluation}

\subsection{Selected LLMs and Agents}
We evaluate ten commercial off-the-shelf LLMs for automated vulnerability repair: GPT-4.1-2025-0414, Gemini-2.5-pro, Doubao1.6-Thinking-0615, Doubao1.6-0615, Deepseek-R1-0528, Deepseek-V3-0324, Kimi-K2, Qwen3-Coder-480b, Qwen3-Max and O3-2025-0416 (see Appendix~\ref{appendix_llm_selection} for details).
For brevity, we omit specific versions of LLMs in the rest of the paper.
In addition to stand-alone LLMs, we also evaluate \update{one program repair agent} that have achieved leading performance on the \update{SWE-Bench~\cite{jimenez2023swe} (i.e., SWE-Agent~\cite{10.5555/3737916.3739517/swe_agent}), one general-purpose code agent for software development (i.e., OpenHands~\cite{wang2025openhands/openhands}), and one commercial vulnerability-repair agent, Claude-Code~\cite{claude_code}.}
Detailed configurations of these agents on \ourbench~are summarized in Appendix~\ref{appendix_agent_config}.

\subsection{Evaluation Settings}
By default, each LLM or agent generates one patch per CVE in a single attempt.
We then measure the effectiveness of the patch through a combination of static matching and dynamic validation, which  aligns with best practices established in recent AVR benchmarks~\cite{nong2025appatchautomatedadaptiveprompting/appatch, PatchAgent, DBLP:conf/issta/WuJPLD0BS23/vjbench, DBLP:journals/corr/abs-2506-11791/secbench}.
All experiments are carried out in isolated sandbox environments to guarantee consistent runtime behaviors across LLMs and agents.

\begin{table}[t]
\scriptsize
\caption{Performance for Vulnerability Repair with Location Oracle}
\begin{adjustbox}{width=0.48\textwidth}
\label{overall}
\setlength{\tabcolsep}{4.8pt} 
\def\arraystretch{1}

\begin{tabular}{
    l 
    >{\raggedleft\arraybackslash}m{0.4cm} 
    >{\raggedleft\arraybackslash}m{0.4cm} 
    >{\raggedleft\arraybackslash}m{0.4cm} 
    >{\raggedleft\arraybackslash}m{0.45cm} |
    >{\raggedleft\arraybackslash}m{0.4cm} 
    >{\raggedleft\arraybackslash}m{0.4cm} 
    >{\raggedleft\arraybackslash}m{0.4cm} 
    >{\raggedleft\arraybackslash}m{0.45cm} |
    >{\raggedleft\arraybackslash}m{0.8cm} 
    >{\raggedleft\arraybackslash}m{0.8cm}
}
\toprule
\multirow{2}{*}{\textbf{\makecell{Evaluated\\ LLMs and Agents}}}  
& \multicolumn{4}{c}{\textbf{Repair (PoC)}}   
& \multicolumn{4}{c}{\textbf{Repair (PoC \& Unit)}}
& \multirow{2}{*}[-0.5ex]{\centering\textbf{\makecell{Avg. \\Token (K)}}}    
& \multirow{2}{*}[-0.5ex]{\centering\textbf{\makecell{Avg.\\Price (\$)}}}              
 \\
\cmidrule(lr){2-5} \cmidrule(lr){6-9} 
& \centering \textbf{Go} & \centering \textbf{Js} & \centering \textbf{Py} & \centering \textbf{Total}
& \centering \textbf{Go} & \centering \textbf{Js} & \centering \textbf{Py} & \centering \textbf{Total}
& & \\ [-0.5ex]
\midrule
\textbf{GPT-4.1}             & \cellcolor[HTML]{E4F0F0}17 & \cellcolor[HTML]{D3E6E6}25 & \cellcolor[HTML]{F5F9F9}9  & \cellcolor[HTML]{9DC7C6}51 & \cellcolor[HTML]{E6F1F1}16 & \cellcolor[HTML]{DCEBEB}21 & \cellcolor[HTML]{FBFDFD}6  & \cellcolor[HTML]{AED0D0}43 & 3.52      & 0.013     \\
\textbf{Gemini-2.5-Pro}      & \cellcolor[HTML]{D8E9E8}23 & \cellcolor[HTML]{D8E9E8}23 & \cellcolor[HTML]{EFF6F6}12 & \cellcolor[HTML]{8EBEBD}58 & \cellcolor[HTML]{DAEAE9}22 & \cellcolor[HTML]{DCEBEB}21 & \cellcolor[HTML]{F5F9F9}9  & \cellcolor[HTML]{9BC6C5}52 & 10.58     & 0.023     \\
\textbf{Doubao1.6-Thinking}  & \cellcolor[HTML]{E0EDED}19 & \cellcolor[HTML]{E6F1F1}16 & \cellcolor[HTML]{FBFDFD}6  & \cellcolor[HTML]{B2D3D2}41 & \cellcolor[HTML]{E2EFEE}18 & \cellcolor[HTML]{EFF6F6}12 & \cellcolor[HTML]{FFFFFF}4  & \cellcolor[HTML]{C1DBDB}34 & 10.28     & 0.017     \\
\textbf{Doubao1.6}           & \cellcolor[HTML]{EDF5F4}13 & \cellcolor[HTML]{D8E9E8}23 & \cellcolor[HTML]{F3F8F8}10 & \cellcolor[HTML]{A7CDCC}46 & \cellcolor[HTML]{EDF5F4}13 & \cellcolor[HTML]{E0EDED}19 & \cellcolor[HTML]{FBFDFD}6  & \cellcolor[HTML]{B8D6D6}38 & 6.89      & 0.009     \\
\textbf{Deepseek-R1}         & \cellcolor[HTML]{E4F0F0}17 & \cellcolor[HTML]{DEECEC}20 & \cellcolor[HTML]{F3F8F8}10 & \cellcolor[HTML]{A5CCCB}47 & \cellcolor[HTML]{E6F1F1}16 & \cellcolor[HTML]{E2EFEE}18 & \cellcolor[HTML]{F3F8F8}10 & \cellcolor[HTML]{ACCFCF}44 & 10.15     & 0.018     \\
\textbf{Deepseek-V3}         & \cellcolor[HTML]{E4F0F0}17 & \cellcolor[HTML]{CDE3E2}28 & \cellcolor[HTML]{F7FBFB}8  & \cellcolor[HTML]{99C4C4}53 & \cellcolor[HTML]{E8F2F2}15 & \cellcolor[HTML]{DCEBEB}21 & \cellcolor[HTML]{FDFEFE}5  & \cellcolor[HTML]{B2D3D2}41 & 3.65      & 0.002     \\
\textbf{Kimi-K2}             & \cellcolor[HTML]{E6F1F1}16 & \cellcolor[HTML]{E0EDED}19 & \cellcolor[HTML]{F7FBFB}8  & \cellcolor[HTML]{AED0D0}43 & \cellcolor[HTML]{E6F1F1}16 & \cellcolor[HTML]{E6F1F1}16 & \cellcolor[HTML]{FBFDFD}6  & \cellcolor[HTML]{B8D6D6}38 & 3.51      & 0.004     \\
\textbf{Qwen3-Coder-480B}    & \cellcolor[HTML]{E2EFEE}18 & \cellcolor[HTML]{D6E7E7}24 & \cellcolor[HTML]{FBFDFD}6  & \cellcolor[HTML]{A3CACA}48 & \cellcolor[HTML]{E4F0F0}17 & \cellcolor[HTML]{E0EDED}19 & \cellcolor[HTML]{FDFEFE}5  & \cellcolor[HTML]{B2D3D2}41 & 3.49      & 0.008     \\
\textbf{Qwen3-Max}            & \cellcolor[HTML]{E6F1F1}16 & \cellcolor[HTML]{D3E6E6}23 & \cellcolor[HTML]{F6FAFA}10  & \cellcolor[HTML]{9EC7C7}49 & \cellcolor[HTML]{E6F1F1}15 & \cellcolor[HTML]{DAEAEA}19 & \cellcolor[HTML]{FBFDFD}7  & \cellcolor[HTML]{AACECE}41 & 3.34      & 0.037     \\
\textbf{O3}                   & \cellcolor[HTML]{DAEAEA}20 & \cellcolor[HTML]{E6F1F1}15 & \cellcolor[HTML]{F9FBFB}7  & \cellcolor[HTML]{A7CDCC}42 & \cellcolor[HTML]{DAEAEA}20 & \cellcolor[HTML]{EBF3F3}13 & \cellcolor[HTML]{F9FBFB}7  & \cellcolor[HTML]{ACCFCF}40 & 5.25      & 0.027     \\
\midrule
\textbf{SWE-Agent-Gemini2.5}  & \cellcolor[HTML]{DAEAEA}20 & \cellcolor[HTML]{C6DEDE}29 & \cellcolor[HTML]{EDF5F5}12 & \cellcolor[HTML]{7BB3B2}61 & \cellcolor[HTML]{DDEBEB}19 & \cellcolor[HTML]{CDE2E2}26 & \cellcolor[HTML]{F6FAFA}8  & \cellcolor[HTML]{8EBEBD}53 & 561.07    & 1.128     \\
\textbf{OpenHands-Gemini2.5}  & \cellcolor[HTML]{DDEBEB}19 & \cellcolor[HTML]{CFE3E3}25 & \cellcolor[HTML]{F6FAFA}8  & \cellcolor[HTML]{90BFBF}52 & \cellcolor[HTML]{DDEBEB}19 & \cellcolor[HTML]{D3E6E6}23 & \cellcolor[HTML]{F9FBFB}7  & \cellcolor[HTML]{97C3C3}49 & 1141.62   & 2.933     \\
\textbf{Claude-Code-Gemini2.5} & \cellcolor[HTML]{DDEBEB}19 & \cellcolor[HTML]{D6E7E7}22 & \cellcolor[HTML]{FBFDFD}6  & \cellcolor[HTML]{9CC6C5}47 & \cellcolor[HTML]{DDEBEB}19 & \cellcolor[HTML]{DDEBEB}19 & \cellcolor[HTML]{FDFEFE}5  & \cellcolor[HTML]{A5CBCB}43 & 333.06   & 0.853   \\

\bottomrule

\end{tabular}
\end{adjustbox}
\end{table}

More specifically, each LLM and agent is evaluated across three dimensions:
(1) \textit{Successful Repair}:
\update{the number of candidate patches that pass security tests (i.e., PoC test) and functionality tests (i.e., unit test).}
(2) \textit{Resource Consumption}:
the average number of tokens consumed, along with the corresponding API costs.
From the cost efficiency perspective, this metric allows comparison between high-performing but expensive approaches versus lightweight alternatives.
(3) \textit{LLM as a Judge}:
To supplement execution-based patch validation, we enlist \update{two independent LLM judges (GPT4.1, and Gemini2.5)} to determine whether a candidate patch is equivalent to the ground truth. 
For clarity, we compare our evaluation settings with those of existing benchmarks in Appendix~\ref{appendix_evaluation_setting}.

\subsection{Research Questions}
\label{sec:RQ}
Armed with \ourbench, we systematically evaluate and compare LLMs and agents on patching real-world vulnerabilities by answering the following research questions (RQs):

\begin{itemize}[leftmargin=12pt]
\item \textbf{RQ1: How accurate are LLMs in generating patches for real-world vulnerabilities?}
This RQ aims to evaluate the performance of LLMs and agents on \ourbench. 
To achieve this goal, we perform experiments with all LLMs and agents under the setting where the location of vulnerable function(s) is provided (S1.1), and with all agents under the setting where vulnerable functions are unknown, referred to as end-to-end repair (S2.2).

\item \textbf{RQ2: How do different prompt strategies affect the performance of LLMs?}
This RQ aims to evaluate the impact of different prompt strategies on LLMs. 
To do so, we perform experiments with all LLMs and agents under six different prompt strategies (as shown in Table~\ref{experiment_setting}): with location oracle (S1.1), without location oracle (S2.1), without Chain-of-Thought (CoT) (S1.2), without external knowledge (S1.3), and with feedback (S1.4 and S2.2).

\item \textbf{RQ3: How do vulnerability localization results affect the performance of LLMs?}
Root cause localization is always critical for vulnerability repair, and this RQ is therefore designed to explore how the localization results influence patch generation.
Specifically, given a vulnerable function, we conduct four sets of experiments, where LLMs are guided with different line-level localization signals, including \textit{Without Line}, \textit{Precise Line}, \textit{Approximate Line}, and \textit{Imprecise Line}, as introduced in Section~\ref{patch_generation_with_location_oracle}.

\item \textbf{RQ4: Do LLMs and agents complement each other in terms of AVR?}
This RQ aims to explore whether LLMs and agents are complementary to each other in AVR. 
To this end, we investigate the number of vulnerabilities that can be uniquely repaired by a single model, and whether different models can repair the same set of vulnerabilities. 

\item \textbf{RQ5: Can LLMs and agents repair vulnerabilities that are hard to fix?}
The difficulty of repairing vulnerabilities varies: some are straightforward to fix, while others are considerably more challenging. 
This RQ aims to investigate whether LLMs and agents are capable of repairing the latter.
A common way to measure the difficulty of vulnerability repair is to gauge the complexity of the patch required to fix a vulnerability~\cite{li2025sok}.
To this end, we categorize vulnerabilities into groups based on the scope of their ground-truth patches --- namely, the number of lines, hunks, and files that must be modified --- and then evaluate how LLMs and agents perform across different groups.

\end{itemize}

\subsection{RQ1: Overall Effectiveness}


\subsubsection{Evaluation with Location Oracle} 
Table~\ref{overall} presents the effectiveness of LLMs and agents in vulnerability repair, with each patch verified through dynamic testing (see Appendix~\ref{llm_as_a_judge_result} for results based on LLM-as-a-judge).
\update{
Specifically, across the evaluated LLMs, Gemini2.5, DeepSeek-R1 and GPT4.1  achieve the best performance, with 52, 44, and 43 successful repairs, respectively.
We observe that PoC failures remain frequent, with LLMs failing to pass PoC tests in 172 to 189 cases, underscoring the challenge of generating security-correct patches.
In addition, unit tests are also a major source of repair failures.
For instance, DeepSeek-V3 produces 12 patches that pass the security tests but fail the corresponding functionality tests, revealing limitations in functional correctness despite vulnerability neutralization.
To further illustrate this issue, we present a case study in~\ref{case_study_agent_LLM_3}.}

\update{\begin{table}[t]
\footnotesize
\caption{Successful Repair of Different LLMs on Agents}
\begin{adjustbox}{width=0.45\textwidth}
\label{differnent_model_in_agent}
\setlength\tabcolsep{3pt}     
\def\arraystretch{0.8}
\begin{tabular}{
    m{1.6cm} 
    >{\raggedleft\arraybackslash}m{1cm} 
    >{\raggedleft\arraybackslash}m{1.8cm} 
    >{\raggedleft\arraybackslash}m{2cm} 
}
\toprule
   & \textbf{GPT4.1} & \textbf{Gemini2.5} & \textbf{Deepseek-R1} \\
\midrule
\textbf{SWE-Agent}   & 49& 53& 45 \\
\textbf{OpenHands}   & 43& 49& 40 \\
\textbf{Claude-Code}   &30 &43&6\\
\bottomrule
\end{tabular}
\end{adjustbox}
\end{table}}

\begin{figure}[t!]
    \centering
    \begin{subfigure}[b]{0.48\textwidth}
        \includegraphics[width=\textwidth]{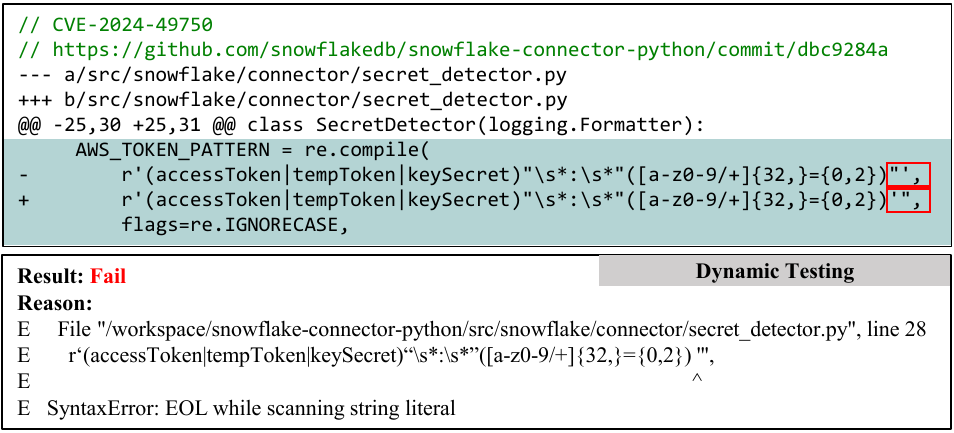}
        \caption{A patch generated by \update{Gemini2.5} for CVE-2024-49750 in \texttt{snowflake-connector-python}. The patch fails to compile due to mismatched quotation marks.}
        \label{case1}
    \end{subfigure}

    \begin{subfigure}[b]{0.48\textwidth}
        \includegraphics[width=\textwidth]{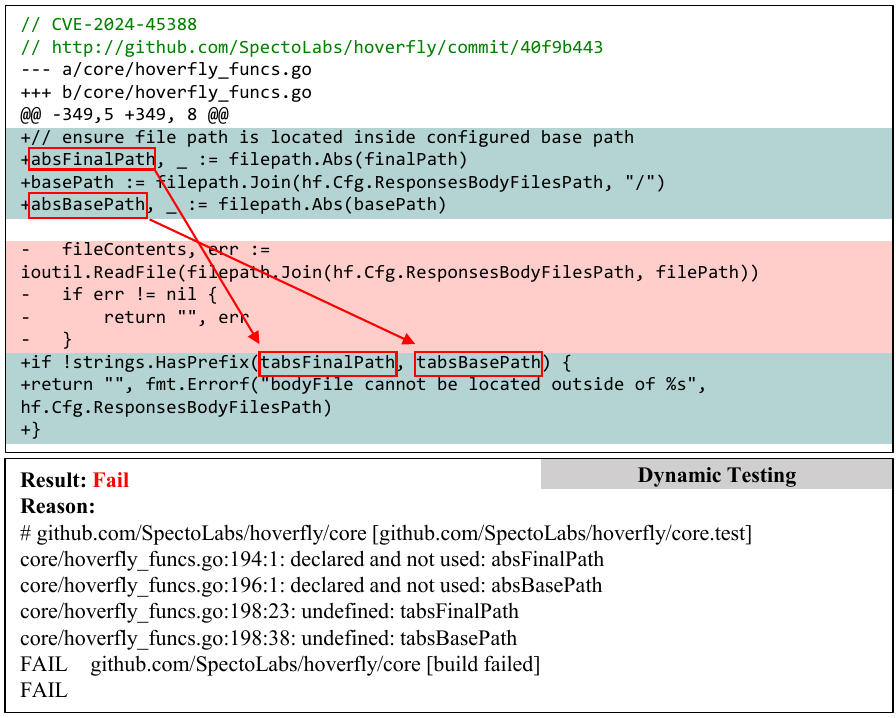}
        \caption{A patch generated by SWE-Agent with \update{Gemini2.5} for CVE-2024-45388 in \texttt{hoverfly}. The patch fails to compile due to the use of undefined variables.}
        \label{case2}
    \end{subfigure}
    \caption{LLM-generated Patches with Compilation Errors}
    \label{case12}
\end{figure}

\update{
Since each agent can be paired with different LLMs, we select the top three models that achieve the best performance in Table~\ref{overall}  (i.e., GPT4.1, Gemini2.5, and DeepSeek-R1) and evaluate their integration with each agent.
As shown in Table~\ref{differnent_model_in_agent}, SWE-Agent generally outperforms standalone LLMs but incurs one to two orders of magnitude higher token consumption (see Table~\ref{overall}).
For example, when paired with SWE-Agent, the number of vulnerabilities repaired by Gemini2.5 increases from 43 to 49, and the number of token consumption increases from 3.52K per repair to 561.07K.
These results suggest that agents, by leveraging external tools and iterative reasoning, can enhance repair effectiveness.
Upon closer inspection, we find that LLMs are particularly prone to syntax-related errors, especially those involving escaping or formatting.
As illustrated in Figure~\ref{case1}, while repairing an information leak vulnerability (CVE-2024-49750), Gemini2.5 produces a patch with mismatched quotation marks, where a double quote follows a single quote, resulting in a compilation failure.
While agents rarely make such mistakes, they remain susceptible to grammar mistakes.
For example, as shown in Figure~\ref{case2}, when SWE-Agent-\update{Gemini2.5} attempts to fix a path traversal vulnerability (CVE-2024-45388), it introduces an undefined variable, which subsequently causes the compiler to raise an undefined variable error.}

\update{Counterintuitively, integrating general-purpose agents does not necessarily lead to performance gains.
In particular, Claude-Code causes a substantial degradation.
For example, when paired with Claude-Code, DeepSeek-R1 achieves only six successful repairs.
An analysis of the 224 failed cases confirms that the agent correctly invokes the LLM; however, 218 of these failures step from timeout exceptions, where the execution of repair exceeds the 30-minute limit.
This result may be attributed to limited cross-model compatibility of Claude-Code --- it is optimized for Claude-series LLMs, and its orchestration may not align well with other models, leading to inefficient tool utilization and prolonged execution cycles.
However, we cannot validate this hypothesis as Claude-Code is closed-source and does not expose intermediate states (e.g., execution trajectories).
It is also worth noting that all three LLMs perform worse when integrated with OpenHands.
This outcome is expected, as OpenHands is primarily designed for general-purpose software development rather than security-oriented vulnerability repair.
In particular, we find that its extensive contextual prompting, while beneficial for general software development, introduces irrelevant context in vulnerability repair, thereby diverting the LLM from patch generation (see Appendix~\ref{case_study_agent_LLM} for a case study).
Overall, these findings underscore the importance of developing domain-specialized agents tailored for AVR, as AVR poses unique challenges that cannot be adequately addressed by general-purpose code agents.}

\update{In summary, the best LLMs repair 43–52 cases, while the program repair agent (SWE-Agent) generally improves performance to 45–53 cases. However, these gains come with substantial token overhead, and general-purpose agents even degrade results, indicating that agent integration is not universally beneficial.}





\begin{table}[t]
\scriptsize
\caption{Performance for End-to-End Vulnerability Repair}
\begin{adjustbox}{width=0.47\textwidth}
\label{end_to_end}
\begin{threeparttable}
\setlength{\tabcolsep}{4.8pt}
\def\arraystretch{1.2}
\begin{tabular}{
    l 
    >{\raggedleft\arraybackslash}m{0.4cm} 
    >{\raggedleft\arraybackslash}m{0.4cm} 
    >{\raggedleft\arraybackslash}m{0.4cm} 
    >{\raggedleft\arraybackslash}m{0.45cm} |
    >{\raggedleft\arraybackslash}m{0.4cm} 
    >{\raggedleft\arraybackslash}m{0.4cm} 
    >{\raggedleft\arraybackslash}m{0.4cm} 
    >{\raggedleft\arraybackslash}m{0.45cm} |
    >{\raggedleft\arraybackslash}m{0.8cm} 
    >{\raggedleft\arraybackslash}m{0.8cm}
}
\toprule
\multirow{2}{*}{\textbf{\makecell{Agents}}}  
& \multicolumn{4}{c}{\textbf{Repair (PoC)}}   
& \multicolumn{4}{c}{\textbf{Repair (PoC \& Unit)}}
& \multirow{2}{*}[-0.5ex]{\centering\textbf{\makecell{Avg. \\Token (K)}}}    
& \multirow{2}{*}[-0.5ex]{\centering\textbf{\makecell{Avg.\\Price (\$)}}}              
 \\ 
\cmidrule(lr){2-5} \cmidrule(lr){6-9} 
& \centering \textbf{Go} & \centering \textbf{Js} & \centering \textbf{Py} & \centering \textbf{Total}
& \centering \textbf{Go} & \centering \textbf{Js} & \centering \textbf{Py} & \centering \textbf{Total}
& & \\ [-0.5ex] \hline
\textbf{SWE-Agent}\tnote{1}                                           & \cellcolor[HTML]{E6F1F0}{13}                     & \cellcolor[HTML]{DBEBEA}{17}                     & \cellcolor[HTML]{F8FBFB}{6}                      & \cellcolor[HTML]{AACECE}{36}                      & \cellcolor[HTML]{E8F2F2}{12}                     & \cellcolor[HTML]{E3EFEF}{14}                     & \cellcolor[HTML]{F8FBFB}{6}                      & \cellcolor[HTML]{B4D4D4}{32}                      & 1190.35                                                                                    & 2.387                                                        \\
\textbf{OpenHands}\tnote{1}                                          & \cellcolor[HTML]{D6E8E8}14 & \cellcolor[HTML]{B6D5D5}25 & \cellcolor[HTML]{EEF5F5}6 & \cellcolor[HTML]{7BB3B2}45 & \cellcolor[HTML]{D6E8E8}14 & \cellcolor[HTML]{B9D7D6}24 & \cellcolor[HTML]{F1F7F7}5 & \cellcolor[HTML]{81B7B6}43 & 1817.37   & 4.640      \\
\textbf{Claude-Code}\tnote{1}                       & \cellcolor[HTML]{EBF4F4}7  & \cellcolor[HTML]{C5DEDD}20 & \cellcolor[HTML]{F4F9F9}4 & \cellcolor[HTML]{A5CBCA}31 & \cellcolor[HTML]{EBF4F4}7  & \cellcolor[HTML]{C8DFDF}19 & \cellcolor[HTML]{F7FAFA}3 & \cellcolor[HTML]{AACFCE}29 & 2492.83   & 6.401     \\

\bottomrule
\multicolumn{11}{l}{\rule{0pt}{1.0em} \textbf{Base LLMs for Agents:} \textsuperscript{1}Gemini2.5}
\end{tabular}
\end{threeparttable}
\end{adjustbox}
\end{table}

\find{{\bf [Finding-1]} With the location oracle, LLMs and agents repair at most \update{23.0\% (53/230) of vulnerabilities }. Specifically, LLMs repair up to \update{52 cases (Gemini2.5)}, whereas agents repair up to \update{53} cases (SWE-Agent with Gemini2.5).}

\subsubsection{End-to-End Repair Evaluation}
To better reflect real-world vulnerability repair scenarios, we also perform end-to-end evaluations without the vulnerable functions provided. 
Specifically, the agent is required to determine the root cause of a vulnerability and its location.
Following prior work~\cite{DBLP:journals/corr/abs-2506-11791/secbench,10.5555/3737916.3739517/swe_agent,wang2025openhands/openhands}, we only evaluate the agents under this setting as shown in Table~\ref{end_to_end}, as standalone LLMs without any navigation tool cannot reliably localize vulnerable code.
\update{OpenHands paired with Gemini2.5 achieves the top repair performance, with 43 successful repairs. }
Compared with Table~\ref{overall}, where vulnerability locations are provided, all agents exhibit a significant decline in performance. 
In addition, as agents need to infer vulnerability locations before repair, they consume more tokens compared with the setting where the locations are provided. 
\update{For example, successful repairs of OpenHands-Gemini2.5 decrease from 49 to 43}, while its average token usage rises from 1,141.62K to 1,817.37K.

\find{{\bf [Finding-2]} In the absence of vulnerability locations, the number of successful repairs declines for all agents (e.g., \update{OpenHands with Gemini2.5: from 49 to 43), accompanied by an increase in token consumption (e.g,. OpenHands with Gemini2.5: from 1,141.62K to 1,817.37K).}}

\begin{table}[t]
\scriptsize
\caption{Impact of Prompt Strategies for Vulnerability Repair with Location Oracle}
\label{ablation_in_llm}
\begin{adjustbox}{width=0.47\textwidth}
\setlength\tabcolsep{8pt}     
\def\arraystretch{0.9}
\begin{tabular}{
    lrrrr
}
\toprule
                     & \textbf{S1.1} & \textbf{S1.2} & \textbf{S1.3} & \textbf{S1.4} \\
\midrule
\textbf{GPT4.1}             & 43 & 43 \textcolor{nicegreen}{($+0$)} & 22 \textcolor{nicered}{($-21$)} & 89~~~\textcolor{nicegreen}{($+46$)}  \\
\textbf{Gemini2.5-Pro}          & 52 & 51 \textcolor{nicered}{($-1$)} & 29 \textcolor{nicered}{($-23$)} & 122~~~\textcolor{nicegreen}{($+70$)} \\
\textbf{Doubao1.6-Thinking} & 34 & 35 \textcolor{nicegreen}{($+1$)} & 21 \textcolor{nicered}{($-13$)} & 81~~~\textcolor{nicegreen}{($+47$)}  \\
\textbf{Doubao1.6}          & 38 & 35 \textcolor{nicered}{($-3$)}  & 26 \textcolor{nicered}{($-12$)}   &  96~~~\textcolor{nicegreen}{($+58$)}   \\
\textbf{Deepseek-R1}        & 44 & 49 \textcolor{nicegreen}{($+5$)} & 30 \textcolor{nicered}{($-14$)} & 105~~~\textcolor{nicegreen}{($+61$)}  \\
\textbf{Deepseek-V3}        & 41 & 43 \textcolor{nicegreen}{($+2$)} & 25 \textcolor{nicered}{($-16$)} & 65~~~\textcolor{nicegreen}{($+24$)}  \\
\textbf{Kimi-K2}            & 38 & 40 \textcolor{nicegreen}{($+2$)} & 23 \textcolor{nicered}{($-15$)} & 89~~~\textcolor{nicegreen}{($+51$)}  \\
\textbf{Qwen3-Coder-480B}   & 41 & 46 \textcolor{nicegreen}{($+5$)} & 25 \textcolor{nicered}{($-16$)} & 78~~~\textcolor{nicegreen}{($+37$)}  \\
\textbf{Qwen3-Max} & 41 & 37 \textcolor{nicered}{($-4$)} & 23 \textcolor{nicered}{($-18$)} & 97~~~\textcolor{nicegreen}{($+56$)} \\ 
\textbf{O3}                 & 40 & 41 \textcolor{nicegreen}{($+1$)} & 28 \textcolor{nicered}{($-12$)} & 113~~~\textcolor{nicegreen}{($+73$)} \\
\midrule
\textbf{SWE-Agent-Gemini2.5}          & 53 & -  & 30 \textcolor{nicered}{($-23$)} & 105~~~\textcolor{nicegreen}{($+52$)}  \\
\textbf{OpenHands-Gemini2.5}   & 49& -      & 48 \textcolor{nicered}{($-~~1$)}  & 129~~~\textcolor{nicegreen}{($+80$)}            \\
\textbf{Claude-Code-Gemini2.5}     &  43 & -   & 37 \textcolor{nicered}{($-~~6$)}  &   -                                                              \\
\bottomrule
\end{tabular}
\end{adjustbox}
\end{table}

\subsection{RQ2: Impact of Prompt Strategies}

Table~\ref{ablation_in_llm} summarizes the performance of LLMs and agents across the four prompt strategies as introduced in Section~\ref{sec:RQ}, which reveals several key observations.

\begin{figure}[t!]
    \includegraphics[width=0.44\textwidth]{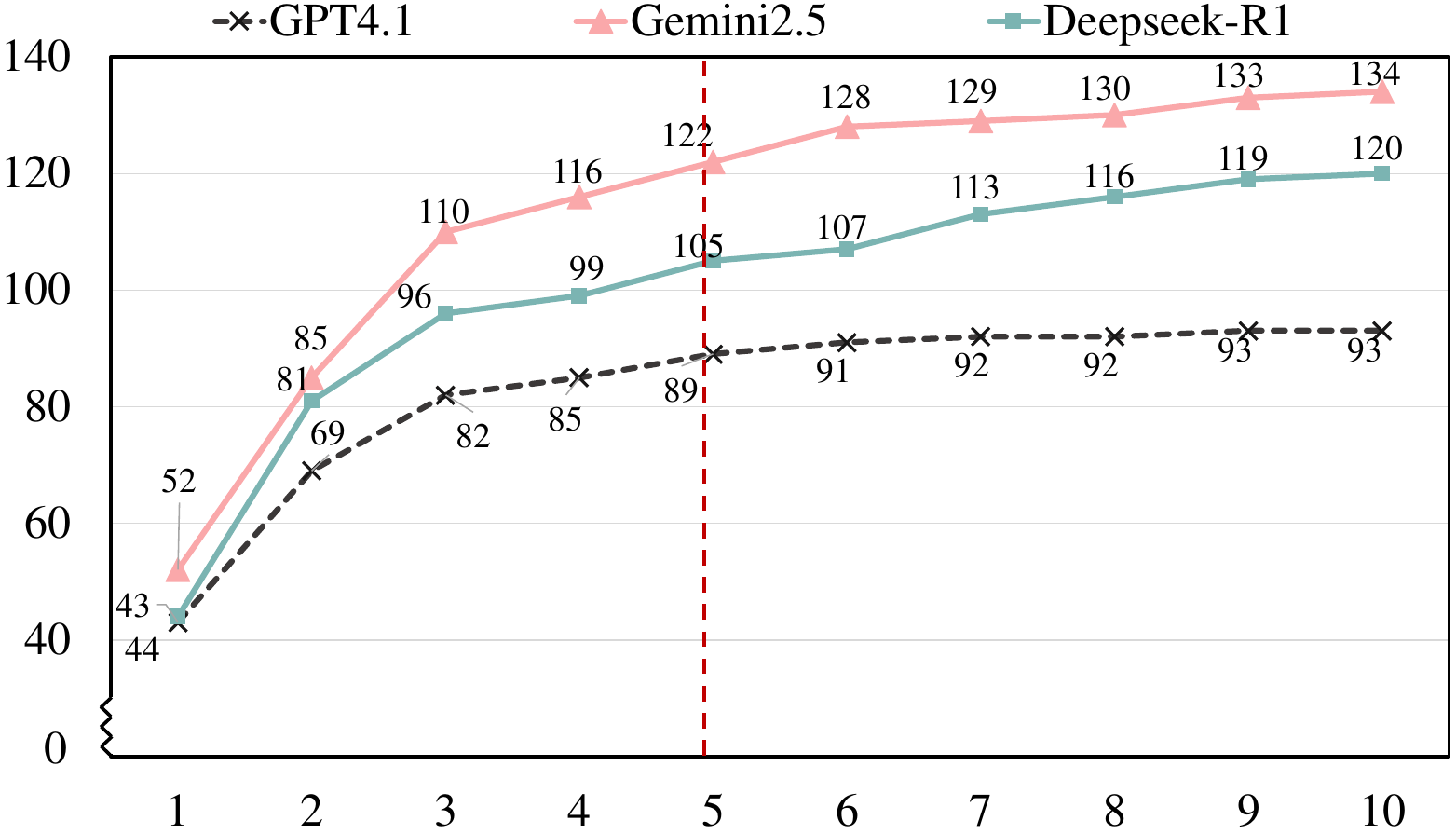}
    \caption{Cumulative Number of Vulnerabilities Solved by LLMs Across Different Rounds of Feedback}
    \label{round of model}
\end{figure}

In terms of LLMs, removing \textit{CoT} (S1.2 vs. S1.1) casts a limited impact: most LLMs only demonstrate marginal differences, suggesting that explicit reasoning contributes little to patch generation with LLM.
A similar result has also been observed in recent AVR research~\cite{san2patch}.  
Second, removing \textit{external knowledge} (S1.3), however, leads to a significant performance decline, especially for GPT4.1, Gemini2.5 and Qwen3-Max, each of which exhibits \update{21, 23 and 18} fewer successful repairs.
This evaluation reveals the importance of domain knowledge in generating valid patches.
Third, \textit{feedback} (S1.4 vs. S1.1) consistently enhances performance, often yielding the highest number of successful repairs across all prompt strategies. 
To further investigate the effect of feedback iterations, we conduct additional experiments on the top three LLMs under varying numbers of feedback rounds.
As shown in Figure~\ref{round of model}, the performance steadily improves as the number of rounds increases, with the optimal results observed at ten rounds.
For example, Gemini2.5 improves from \update{52 in S1.1 to 134} successful repairs after ten rounds. 
This observation highlights the importance of iterative patch validation in improving the quality of patches.
Nevertheless, Figure~\ref{round of model} indicates diminishing returns, as performance gains plateau after roughly five rounds. 
Accordingly, considering the trade-off between cost, efficiency, and performance, we select five rounds of feedback as the default setting for our evaluations in S1.4.

\begin{table}[t]
\footnotesize
\centering
\setlength\tabcolsep{8pt}     
\def\arraystretch{0.9}
\caption{Impact of Prompt Strategies for Agents Without Location Provided}
\label{ablation_in_agent}
\begin{adjustbox}{width=0.47\textwidth}
\begin{tabular}{
    l
    >{\raggedleft\arraybackslash}m{1cm} 
    >{\raggedleft\arraybackslash}m{1cm} 
    >{\raggedleft\arraybackslash}m{1cm} 
}
\toprule
                            \textbf{Agents}   & \textbf{S1.1} & \textbf{S2.1} & \textbf{S2.2}  \\
\midrule
\textbf{SWE-Agent-Gemini2.5} &53 & 32 &87 \\
\textbf{OpenHands-Gemini2.5} &49 & 43 &82 \\
\textbf{Claude-Code-Gemini2.5} &43 &29 &-\\

\bottomrule
\end{tabular}
\end{adjustbox}
\end{table}

As for agents, as shown in Table~\ref{ablation_in_agent}, removing the location of vulnerable functions (S1.1 vs. S2.1) results in a performance decline across all agents. 
For example, SWE-Agent with Gemini2.5 drops \update{from 53 to 32} repairs, while \update{Claude-Code with Gemini2.5 decreases from 43 to 29}, underscoring the critical role of vulnerability localization in guiding patch generation.
In contrast, introducing feedback (S2.1 vs. S2.2) substantially enhances the performance of end-to-end vulnerability repair.
For instance,  \update{SWE-Agent with Gemini2.5 improves from 32 to 87, and OpenHands with Gemini2.5 rises from 43 to 82}.

Overall, we find that accurate vulnerability localization lays the groundwork for repair, while feedback-driven reflection is key for LLMs to achieve stronger performance.

\find{{\bf [Finding-3]} In terms of prompt strategies, feedback delivers the largest gains \update{(\eg~SWE-Agent with Gemini2.5: from 32 to 87),} removing vulnerability knowledge reduces the number of successful repairs by up to \update{23}, and CoT provides minimal difference.}




\begin{table}[t]
\footnotesize
\caption{Impact of Vulnerability Localization Precision}
\label{patch_location}
\begin{adjustbox}{width=0.47\textwidth}
\setlength\tabcolsep{6pt}     
\def\arraystretch{0.9}
\begin{tabular}{
 lcccc}
\toprule
\textbf{Models} & \textbf{Without} & \textbf{Precise} & \textbf{Approximate} & \textbf{Imprecise} \\
\midrule
\textbf{GPT4.1} & 43 & 50 & 41 & 36  \\
\textbf{Gemini2.5}  & 52 & 62 & 51 & 49  \\
\textbf{Deepseek-R1}  & 44 & 51 & 49 & 43  \\ 
\midrule
\textbf{Average} & 46.3 & \textbf{54.3} & 47.0 & 42.7 \\
\bottomrule
\end{tabular}
\end{adjustbox}
\end{table}

\subsection{RQ3: Impact of Localization Results} 
Table~\ref{patch_location} summarizes the performance of the top three LLMs under different vulnerability localization settings, as introduced in Section~\ref{sec:RQ}.
Note that this experiment is conducted under the default setting S1.1. 
The results reveal a clear trend: precise vulnerable locations consistently yield the best performance.
\update{For example, Gemini2.5 improves from 52 repairs (Without Line) to 62 (Precise Line).}
This is within our expectations as accurate localization reduces the search space for LLMs to fix vulnerabilities.
In contrast, inaccurate localization may mislead LLMs in identifying the root cause of the vulnerability.
\update{Case in point, the Imprecise Line setting performs worse than the baseline (Without Line), with the average number of successful repairs decreasing \update{from 46.3 to 42.7}.}
Overall, vulnerability localization casts contrasting effects: accurate localization can substantially enhance repair effectiveness, whereas inaccurate guidance can hinder it.

\find{{\bf [Finding-4]} Accurate vulnerability locations enhance repair success (average rising \update{from 46.3 to 54,3}), whereas inaccurate locations degrade performance (average dropping \update{from 46.3 to 42.7}), underscoring the importance of vulnerability localization in repair.}

\begin{figure}[t!]
    \centering
    \begin{subfigure}{0.22\textwidth}
        \centering
        \includegraphics[width=1\textwidth]{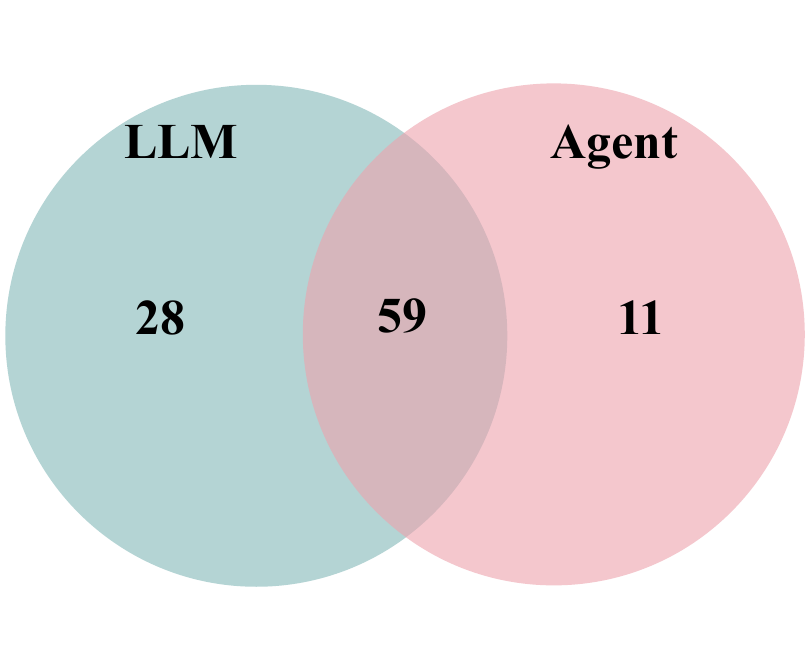} 
        \caption{Interaction of LLMs and Agents}
        \label{venn1}
    \end{subfigure}
    \begin{subfigure}{0.205\textwidth}
        \centering
        \includegraphics[width=1\textwidth]
        {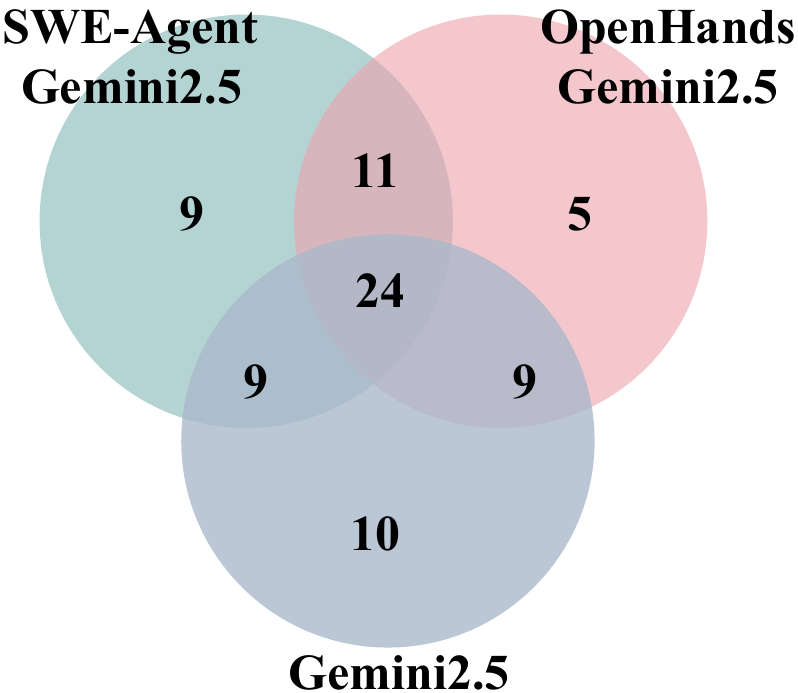}
        
        \caption{Interaction of the Top Three Approaches}
        \label{venn2}
    \end{subfigure}
    \caption{Interactions of Successful Repairs Across Models}
    \label{venn_llm_agent}
\end{figure} 
\begin{figure}[t!]
    \includegraphics[width=0.45\textwidth]{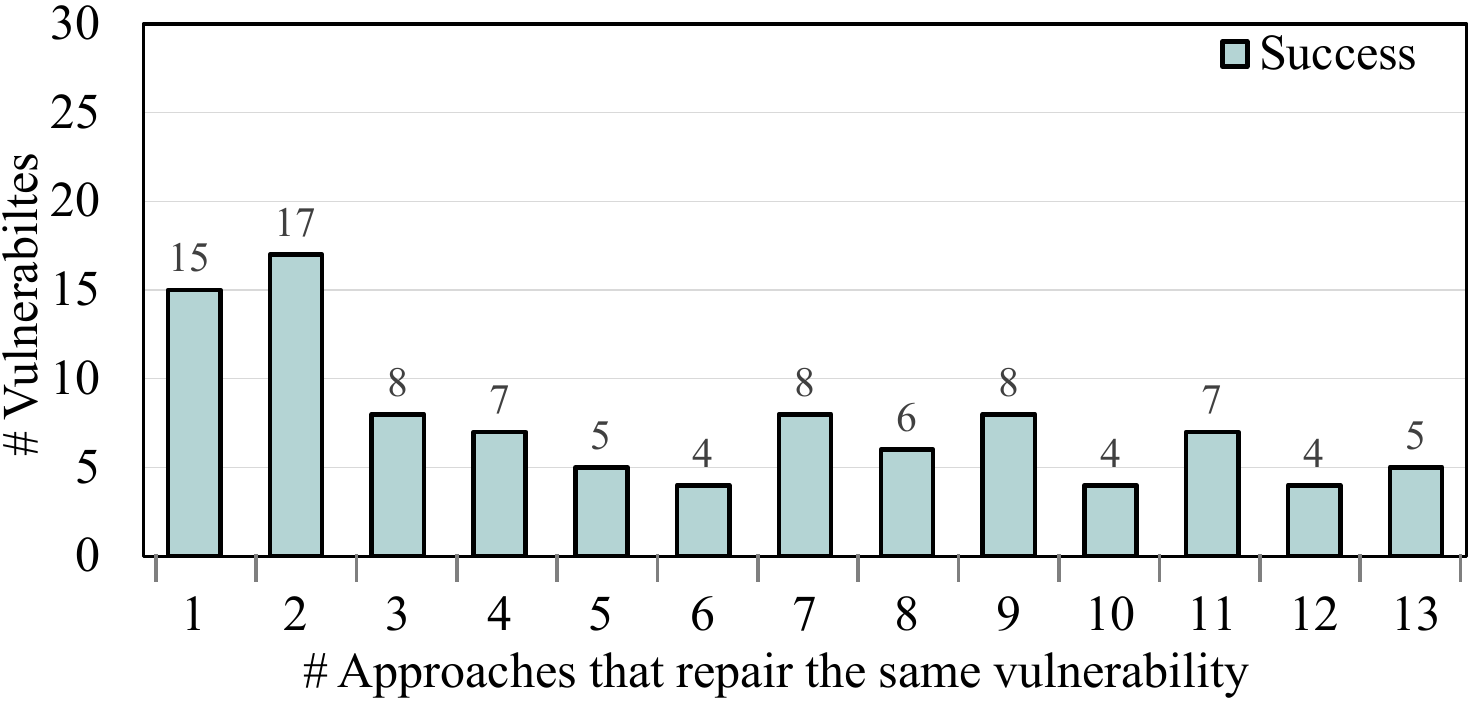}
    \caption{Overlapping Vulnerability Repairs}
    \label{temp}
\end{figure}

\subsection{RQ4: Complementary of LLMs/Agents}
In this RQ, we investigate if the same vulnerability can be repaired by different approaches (in total 13 as listed in Table~\ref{overall}).
Figure~\ref{temp} summarizes the results. 
Overall, \update{98} CVEs can be repaired by at least one LLM or agent. 
Compared to the \update{53} CVEs repaired by the best-performing approach (Claude-Code with \update{Gemini2.5}), this leaves a substantial gap of \update{45}, highlighting considerable room for improvement.
In contrast, only \update{five} cases are consistently repaired by all approaches, whereas \update{15} cases are uniquely addressed by a single approach. 
Furthermore, we observe that most vulnerabilities are fixed by only a limited subset of approaches. 
For example, \update{52} cases are repaired by five or fewer approaches.

\begin{table*}[t]
\centering
\setlength\tabcolsep{8pt}     
\def\arraystretch{0.8}
\caption{Performance of LLMs for Vulnerability Repair across Various Levels of Patch Complexities}
\begin{adjustbox}{width=0.95\textwidth}
\label{performance-cve-complexity}
\scriptsize
\begin{tabular}{l
rrrrr|
rrrrrr|
rrrr}
\toprule
\textbf{}                   & \multicolumn{5}{c|}{\textbf{Line}}         & \multicolumn{6}{c|}{\textbf{Hunk}}    & \multicolumn{4}{c}{\textbf{File}} \\
\cmidrule(lr){2-6} \cmidrule(lr){7-12} \cmidrule(lr){13-16} 

\textbf{}                   & \textbf{1-5} & \textbf{5-10} & \textbf{10-20} & \textbf{20-50} & \textbf{50+} & \textbf{1}  & \textbf{2}  & \textbf{3}  & \textbf{4}  & \textbf{5} & \textbf{6+} & \textbf{1}     & \textbf{2}    & \textbf{3}   & \textbf{4+}  \\
\midrule
\textbf{GPT4.1}               & 17            & 10            & 10           & 6            & 0            & 28            & 14            & 0            & 1            & 0            & 0   & 42   & 1   & 0   & 0   \\
\textbf{Gemini2.5}            & 22            & 14            & 10           & 4            & 2            & 36            & 12            & 1            & 1            & 2            & 0   & 48   & 4   & 0   & 0   \\
\textbf{Doubao1.6-Thinking}   & 17            & 10            & 5            & 2            & 0            & 26            & 8             & 0            & 0            & 0            & 0   & 33   & 1   & 0   & 0   \\
\textbf{Doubao1.6}            & 20            & 7             & 6            & 5            & 0            & 28            & 10            & 0            & 0            & 0            & 0   & 37   & 1   & 0   & 0   \\
\textbf{Deepseek-R1}          & 18            & 12            & 9            & 5            & 0            & 30            & 12            & 1            & 1            & 0            & 0   & 43   & 1   & 0   & 0   \\
\textbf{Deepseek-V3}          & 16            & 12            & 6            & 7            & 0            & 29            & 11            & 1            & 0            & 0            & 0   & 40   & 1   & 0   & 0   \\
\textbf{Kimi-K2}              & 21            & 9             & 4            & 4            & 0            & 27            & 10            & 0            & 1            & 0            & 0   & 38   & 0   & 0   & 0   \\
\textbf{Qwen3-Coder-480B}     & 18            & 12            & 8            & 3            & 0            & 31            & 9             & 0            & 1            & 0            & 0   & 41   & 0   & 0   & 0   \\
\textbf{Qwen3-Max}            & 16            & 11            & 11           & 3            & 0            & 27            & 12            & 1            & 1            & 0            & 0   & 40   & 1   & 0   & 0   \\
\textbf{O3}                   & 20            & 9             & 6            & 4            & 1            & 28            & 11            & 0            & 0            & 1            & 0   & 39   & 1   & 0   & 0   \\
\midrule
\textbf{SWE-Agent-Gemini2.5}  & 23            & 13            & 11           & 6            & 0            & 39            & 11            & 1            & 2            & 0            & 0   & 50   & 3   & 0   & 0   \\
\textbf{OpenHands-Gemini2.5}  & 25            & 7             & 11           & 5            & 1            & 35            & 10            & 2            & 1            & 1            & 0   & 48   & 1   & 0   & 0   \\
\textbf{Claude-Code-Gemini2.5} & 16            & 9             & 11           & 6            & 1            & 31            & 8             & 1            & 2            & 1            & 0   & 41   & 2   & 0   & 0   \\
\midrule
\textbf{Repaied on Average} & \textbf{19.2} & \textbf{10.4} & \textbf{8.3} & \textbf{4.6} & \textbf{0.4} & \textbf{30.4} & \textbf{10.6} & \textbf{0.6} & \textbf{0.8} & \textbf{0.4} & \textbf{0.0} & \textbf{41.5} & \textbf{1.3} & \textbf{0.0} & \textbf{0.0} \\
\textbf{Total Number in Group}             & \textbf{59}   & \textbf{44}   & \textbf{62}   & \textbf{46}  & \textbf{19}  & \textbf{120}  & \textbf{69}   & \textbf{17}  & \textbf{12}  & \textbf{7}   & \textbf{5}   & \textbf{202}  & \textbf{21}  & \textbf{6}   & \textbf{1}   \\
\bottomrule
\end{tabular}
\end{adjustbox}
\end{table*}

To gain deeper insights, we analyze the overlap of correctly repaired CVEs across different approaches.
Specifically, we conduct two comparisons. 
\update{First, we compare the repair results of ten LLMs and three agents.} 
Second, we study the overlap among the top three approaches:  \update{SWE-Agent with Gemini2.5}, OpenHands with \update{Gemini2.5}, and \update{Gemini2.5}.
Figure~\ref{venn_llm_agent} presents the Venn diagrams that illustrate the overlap of successful repairs across the selected approaches. 
Figure~\ref{venn1} shows that the union of all approaches yields  correctly repaired vulnerabilities.
It is worth noting that each approach exhibits distinct advantages, with substantial numbers of unique repairs (\eg~\update{ten} for \update{Gemini2.5 and nine for SWE-Agent with Gemini2.5}).
These results highlight the strong complementarity among existing models and underscore the need for future research to integrate these complementary capabilities, thereby achieving higher overall repair performance.

\find{{\bf [Finding-5]} Models exhibit strong complementarity in vulnerability repair, often producing distinct sets of successful patches. Specifically, \update{98} vulnerabilities can be repaired by at least one model, yet only \update{five} are fixed by all models; \update{15} are uniquely addressed by a single model.}

\subsection{RQ5: Effectiveness on Complex Cases}
\label{RQ5}
Table~\ref{performance-cve-complexity} reports the effectiveness of LLMs and agents across different complexities of patch oracle.
Specifically, we define the complexity of a patch as the number of lines, hunks, and files to be modified for vulnerability repair.
For patched lines, we divide vulnerabilities into five groups: 1–5, 5–10, 10–20, 20–50, and more than 50 lines.
For patch hunks, vulnerabilities are divided into six groups: five groups with 1–5 hunks each, and one group with over five hunks.
For patched files, vulnerabilities are categorized into four groups based on the number of modified files.

Experimental results show that most successful repairs are concentrated in relatively simple patches, such as those involving fewer than 10 line modifications or only one to two hunks. 
For example, all methods successfully repair \update{32.5\% (19.2/59)} vulnerabilities on average in the 1–5 line group and \update{25.3\% (30.4/120)}  in the single-hunk group.
In contrast, repair performance declines as patch complexity increases.
When patch modifications grow to 5–10 lines or two hunks, the repair rates on average drop to \update{23.6\% (10.4/44)} and \update{15.4\% (10.6/69)}. 
More importantly, current approaches rarely repair highly complex vulnerabilities. 
For illustration, we provide a case study in Appendix~\ref{case_study_agent_LLM_2}.
On average, the repair rates fall dramatically: \update{7.7\% (5.5/65)} for patches involving over 20 lines, \update{5.0\% (1.2/24)} for patches with over three hunks, and \update{0\% (0.0/7)} for patches spanning over two files.

Overall, current approaches are more effective at generating relatively simple vulnerability patches.
Success repair rates, however, drop substantially for more complex patches, suggesting the need for more advanced techniques that can reason over larger code contexts.

\find{{\bf [Finding-6]} Most successful repairs occur in less complex patches. For example, the average repair rate for simple CVEs (patches modifying 1–5 lines) is \update{32.5\%}, whereas it drops to \update{7.7\%} for complex CVEs (patches modifying 20+ lines), underscoring the difficulty of repairing vulnerabilities that require extensive code changes.}




\section{Threat to Validity}
There are four main threats to the validity of this study.
First, \ourbench~focuses on three programming languages, i.e., Python, JavaScript, and Go. 
They are chosen because they are widely used, frequently reported in CVEs, and underexplored in prior AVR studies. 
However, the exclusion of other languages may limit the generalization of our findings.
Second, a patch may pass all security tests and unit tests without truly fixing the vulnerability due to the limited coverage of test suites.
To mitigate this risk, we manually review the test cases to ensure that they exercise the relevant vulnerable code paths.
Nonetheless, manual examination cannot guarantee completeness, leaving residual risk in patch validation.
Third, since LLMs are trained on massive corpora, it is difficult to rule out the possibility that CVEs or their patches in \ourbench~are included in LLM's pretraining data. 
Such overlap could result in data leakage and artificially inflated performance.
\update{Finally, the inherent randomness during LLM inference can cause variations in vulnerability repair, resulting in different patches across runs. 
To minimize such variability, we set the temperature parameter to 0 during inference whenever possible, aiming to yield more consistent patch outputs.}

\section{Conclusion}
In this paper, we introduce \ourbench, a comprehensive benchmark for vulnerability repair that consists of 1,000 real-world CVEs across 65 CWE categories.
Through an evaluation of ten LLMs and four agents on \ourbench, we find that the best-performing LLM achieves a repair rate of only \update{22.6\%} and the best-performing agent reaches \update{23.0\%}, revealing a significant gap for real-world deployment.
Our analysis further underscores the importance of vulnerability localization and patch validation in effective vulnerability repair.
We also notice that existing models are highly complementary to each other, highlighting the need for future research to integrate various models' strengths, thereby
achieving higher repair performance. 
We believe the presented benchmark together with our findings are beneficial for future AVR research. 

\section*{Ethical Considerations}
Our work strictly adheres to responsible security research practices and ethical guidelines. 
All vulnerabilities in \ourbench{} are drawn from publicly disclosed CVEs, and all corresponding patches are collected from open-source GitHub repositories, ensuring no exposure of undisclosed or proprietary code. 
To prevent potential harm, all experiments are conducted on a private server within isolated, dockerized environments, eliminating risks of unintended interactions with external systems or real-world users.
\ourbench{} is designed as a benchmark for evaluating LLMs in automated vulnerability repair. 
Since every included vulnerability has already been patched in upstream projects, the dataset cannot be used to mount real-world attacks. 
Instead, our benchmark advances transparent and reproducible evaluation of automated vulnerability repair methods, thereby contributing to the broader ethical goal of strengthening software security.

\section*{Open Science}
We fully support the principles of open science, emphasizing transparency and reproducibility. To this end, all artifacts generated in this work are released at \url{https://github.com/bytedance/PatchEval}, including the benchmark dataset, sandbox environments, evaluation framework, experiment logs, and documentation needed to reproduce the main results reported in the paper.
By making these artifacts publicly available, we enable researchers to validate our findings, reproduce our experiments, and benchmark new approaches.

\section*{Acknowledgment}
We would like to thank Caiyong Lin, Guangyu Zhou, Sen Cheng, Xufeng Zhou, Ke Sun, Jinhuang Liang, Zhongfu Su, Pengfei Sun, Zequn Fang, and Yongheng Yang at ByteDance for their dedicated efforts in reviewing the quality of the dataset. This work would not have been possible without their support.
We thank Zhengqin Luo, Zhi Liu, Zach Zhang, and Yuan Zhang for their valuable feedback and advice.
We also thank Shengqiang Li for helping artifact evaluation.
Any opinions, findings, and conclusions or recommendations expressed in this work are those of the authors and do not necessarily reflect the views of any institution.
 
{\footnotesize \bibliographystyle{acm}
\bibliography{sample}}
\appendix
\section{Appendix}

\subsection{LLMs Selected for AVR Evaluation}
\label{appendix_llm_selection}
\begin{table}[htbp]
\begin{adjustbox}{width=0.45\textwidth}
\centering
\begin{threeparttable}
\footnotesize
\def\arraystretch{0.9} 
\caption{Ten LLMs Selected for AVR Evaluation}
\label{LLM_selection}
\scriptsize 
\begin{tabular}{lrrll}
\toprule
\textbf{Models}              & \textbf{Params} & \textbf{Max Token} & \textbf{Release} & \textbf{Source}                                                                            \\
\midrule
\textbf{GPT-4.1-2025-0414}  & NA\tnote{1}               & 1000K              & 2025-4-14              & \href{https://platform.openai.com/docs/models/gpt-4.1}{OpenAI}                                                  \\
\textbf{Gemini-2.5-Pro}     & NA               & 1000K              & 2025-6-17              & \href{https://cloud.google.com/vertex-ai/generative-ai/docs/models/gemini/2-5-pro?hl=zh-cn}{Google}             \\ 
\textbf{Doubao1.6-Thinking-0615}  & NA               & 256K               & 2025-6-11              & \href{https://www.volcengine.com/docs/82379/1593703}{ByteDance}                                                    \\
\textbf{Doubao1.6-0615}          & NA               & 256K               & 2025-6-11              & \href{https://www.volcengine.com/docs/82379/1593702}{ByteDance}                                                    \\
\textbf{Deepseek-R1-0528}   & 671B            & 128K               & 2025-5-28              & \href{https://api-docs.deepseek.com/zh-cn/quick_start/pricing}{Deepseek}                                                   \\
\textbf{Deepseek-V3-0324}   & 671B            & 128K               & 2025-3-24              & \href{https://api-docs.deepseek.com/zh-cn/quick_start/pricing}{Deepseek}                                                    \\ 
\textbf{Kimi-K2}            & 1000B           & 128K               & 2025-7-11              & \href{https://platform.moonshot.cn/docs/pricing/chat}{MoonShot} \\
\textbf{Qwen3-Coder-480B}   & 480B            & 256K               & 2025-7-22              & \href{https://huggingface.co/Qwen/Qwen3-Coder-480B-A35B-Instruct}{Alibaba}                                       \\
\textbf{Qwen3-Max}   & NA            & 256K               & 2025-9-23              & \href{https://qwen.ai/blog?id=241398b9cd6353de490b0f82806c7848c5d2777d&from=research.latest-advancements-list}{Alibaba}                                       \\
\textbf{O3-2025-0416}       & NA               & 200K               & 2025-4-16              & \href{https://platform.openai.com/docs/models/o3}{OpenAI}   \\ 
\bottomrule
\multicolumn{5}{l}{\textsuperscript{1}Not publicly accessible.} \\
\end{tabular}
\end{threeparttable}  
\end{adjustbox}
\end{table}
Table~\ref{LLM_selection} presents the ten LLMs we select to evaluate for their AVR capabilities. 
To obtain deterministic and reproducible results, we set the temperature parameter to 0 for LLMs that support this option, while retaining default configurations for those that do not.

\subsection{Agent Configurations on \ourbench}
\label{appendix_agent_config}
\update{For SWE-Agent and OpenHands}, we replicate their default workflows for program repair and software development, adaptiing them to our vulnerability repair settings.
For Claude-Code, we design an automated workflow compatible with \ourbench{} by emulating its built-in security review functionality.
\update{To ensure consistency across agents, we apply a standardized configuration protocol: each agent is restricted to a maximum of 100 interactions or 30-minute runtime within the sandbox to control execution time and resource usage.}
During patch generation, we revise the default prompts by adding CVE-specific details so that the generated patches are grounded in the vulnerability contexts.
For patch validation, we implement a customized PoC framework that executes security and functionality tests, returns validation results, and supports iterative agent interactions.
It is worth noting that none of the agents perform S1.2, as their thought–action phase inherently follows a chain-of-thought process.
Moreover, because Claude-Code is closed-source, we cannot integrate additional tools for security testing. 
Consequently, Claude-Code does not support S1.4 and S2.2 in our experiments, as both settings require security testing to enable \textit{feedback}.

\subsection{Evaluation Settings in Existing Benchmarks}
\label{appendix_evaluation_setting}
\begin{table*}[t]
\centering
\caption{Evaluation Settings in Existing Benchmarks}
\label{compare_benchmark_experimental_setting}
\begin{adjustbox}{width=0.95\textwidth}
\begin{threeparttable}
\footnotesize
\def\arraystretch{0.5}
\begin{tabular}{l cc | ccc | cc | ccc}
\toprule

  & \multicolumn{2}{c}{\textbf{Task Formulation}} 
  & \multicolumn{3}{c}{\textbf{Evaluation Metric}} 
  & \multicolumn{2}{c}{\textbf{Evaluated Approach}} 
  & \multicolumn{3}{c}{\textbf{Prompt Strategy}} \\
\cmidrule(lr){2-3} \cmidrule(lr){4-6} \cmidrule(lr){7-8} \cmidrule(lr){9-11}
  
  & \textbf{w/ Location Oracle} & \textbf{End to End} 
  & \textbf{Static} & \textbf{Dynamic} & \textbf{Cost} 
  & \textbf{LLM} & \textbf{Agent} 
  & \textbf{CoT} & \textbf{Knowledge} & \textbf{Feedback} \\
\midrule
Big-Vul~\cite{DBLP:conf/msr/FanL0N20/bigvul}     &\ding{55}  &\ding{55}   &\ding{55}   &\emptydiamond{}\tnote{1}   &\ding{55}  &\emptycirc   &\emptycirc  &\ding{55}   &\ding{55}   &\ding{55}   \\
ExtractFix~\cite{DBLP:journals/tosem/GaoWDJXR21/extractfix}  &\ding{55}  &\ding{55}   &\ding{55}   &\emptydiamond{}   &\ding{55}   &\emptycirc   &\emptycirc &\ding{55}   &\ding{55}   &\ding{55}   \\
CVEfixes~\cite{DBLP:conf/promise/BhandariNM21/cvefixes}    &\ding{55}  &\ding{55}   &\ding{55}   &\emptydiamond{}   &\ding{55}   &\emptycirc   &\emptycirc  &\ding{55}   &\ding{55}   &\ding{55}   \\
PatchDB~\cite{DBLP:conf/dsn/WangWF0J21/patchdb}    &\ding{55}  &\ding{55}   &\ding{55}   &\emptydiamond{}   &\ding{55}   &\emptycirc   &\emptycirc  &\ding{55}   &\ding{55}   &\ding{55}   \\

VJBench~\cite{DBLP:conf/issta/WuJPLD0BS23/vjbench} & \ding{51}  & \ding{55} & \ding{51} &\fulldiamond{}\tnote{2}  & \ding{55} & \fullcirc &  \emptycirc\tnote{4} & \ding{55}  &  \ding{55} &\ding{55}          \\
Sec-Bench~\cite{DBLP:journals/corr/abs-2506-11791/secbench} & \ding{55} & \ding{51} & \ding{55}  &\halfdiamond{}  & \ding{51}  &   \emptycirc  & \fullcirc\tnote{5} &\ding{55} &  \ding{55} & \ding{55}         \\
CoV-Eval~\cite{DBLP:journals/corr/abs-2505-10494/CoV-Eval}  & \ding{51} &  \ding{55} & \ding{51}  &\emptydiamond{}  & \ding{55} & \fullcirc &  \emptycirc & \ding{55}  & \ding{51} &\ding{55}          \\
ARVO~\cite{DBLP:journals/corr/abs-2408-02153/arvo}  & \ding{51}  &  \ding{55} & \ding{55} &\halfdiamond{}  &  \ding{55}  & \fullcirc & \emptycirc  &\ding{55} & \ding{51} & \ding{55}         \\
VADER~\cite{DBLP:journals/corr/abs-2505-19395/vader} & \ding{55} & \ding{51} & \ding{55} &\emptydiamond{}  &   \ding{55} & \fullcirc  & \emptycirc  & \ding{51} & \ding{51} &  \ding{55}        \\
Vul4C~\cite{DBLP:journals/corr/abs-2506-11697/vul4c}  &\ding{55}  &\ding{55}   &\ding{55}   &\fulldiamond{}   &\ding{55}   &\emptycirc   &\emptycirc  &\ding{55}   &\ding{55}   &\ding{55}   \\
CVE-Bench~\cite{DBLP:conf/naacl/WangLX25a/cvebench} & \ding{55} & \ding{51} & \ding{55}  &\halfdiamond{}  & \ding{55} & \emptycirc & \halfcirc{}\tnote{6}  & \ding{51} & \ding{55}  & \ding{55}       \\
\midrule
\textbf{\ourbench}  & \ding{51} & \ding{51}  & \ding{51} & \fulldiamond{}\tnote{3}  &   \ding{51} & \fullcirc & \fullcirc & \ding{51} & \ding{51} & \ding{51}       \\
\bottomrule
\multicolumn{11}{l}{
\textsuperscript{1}~\update{\emptydiamond No security or functionality test.}
\textsuperscript{2}~\update{\halfdiamond Only security test.}
\textsuperscript{3}~\update{\fulldiamond Both security and functionality test.}
\textsuperscript{4}~\emptycirc No evaluation method.
\textsuperscript{5}~\halfcirc Only one evaluation method.
}\\
\multicolumn{11}{l}{
\textsuperscript{6}~\fullcirc More than one evaluation methods.
}
\\
\end{tabular}
\end{threeparttable}
\end{adjustbox}
\end{table*}
Table~\ref{compare_benchmark_experimental_setting} compares the evaluation settings between \ourbench~with existing AVR benchmarks.


\subsection{Evaluation by LLM-as-a-Judge}
\label{llm_as_a_judge_result}
\begin{table*}[!t]
\centering

\caption{Performance of Top-3 LLMs (Patch Evaluation with LLM-as-a-Judge)}
\label{llm_judge_result_table}
\begin{adjustbox}{width=0.65\textwidth}
\begin{threeparttable}
\def\arraystretch{0.9}
\small
\begin{tabular}{
 l
 ccc |
 ccc
}
\toprule

 \textbf{}  & \multicolumn{3}{c}{\textbf{GPT4.1-Judge}} & \multicolumn{3}{c}{\textbf{Gemini2.5-Judge}} \\
\cmidrule(lr){2-4} \cmidrule(lr){5-7} 
\textbf{}& \textbf{PoC}\tnote{1} & \textbf{Score}\tnote{2} & \multicolumn{1}{c}{\textbf{Equivalence}\tnote{3}}  & 
\textbf{PoC} & \textbf{Score}  & \textbf{Equivalence}  \\
\midrule
\textbf{GPT4.1} & 32 & 81.14\% & 160 & 47 & 77.19\% & 254 \\
\textbf{Gemini2.5}     & 30 & 76.32\% & 160 & 39 & 77.63\% & 239 \\
\textbf{Deepseek-R1}   & 31 & 80.35\% & 155 & 45 & 79.48\% & 231\\
\bottomrule
\end{tabular}
\begin{tablenotes}
\footnotesize
\item[1]~PoC Validation Dataset
\item[2]~The consistency between LLM-judge results and dynamic testing results
\item[3]~Results on Equivalence Validation Dataset with 1000 CVEs
\end{tablenotes}   
\end{threeparttable}
\end{adjustbox}
\end{table*}


\begin{figure}[t!]
    \centering
    \begin{subfigure}[b]{0.48\textwidth}
        \includegraphics[width=\textwidth]{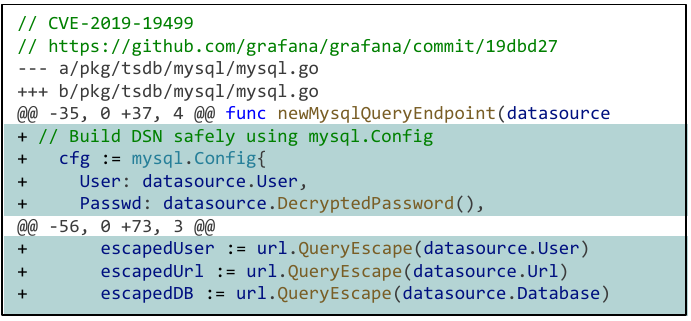}
        \caption{Patch for CVE-2019-19499 in \texttt{grafana}}
        \label{llm_judge_case_patch}
    \end{subfigure}

    \begin{subfigure}[b]{0.48\textwidth}
        \includegraphics[width=\textwidth]{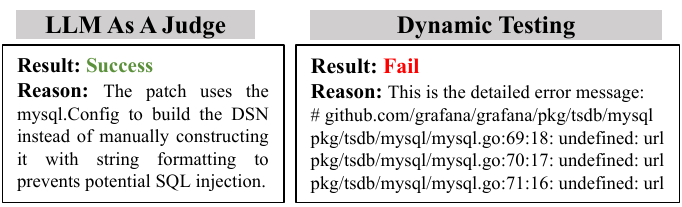}
        \caption{Patch Verification Results by Different Methods}
        \label{llm_judge_case_result}
    \end{subfigure}
    \caption{Patch Verification for CVE-2019-19499}
    \label{llm_judge_case}
\end{figure}
Table~\ref{llm_judge_result_table} summarizes the effectiveness of the three best-performing LLMs in Table~\ref{overall}.
Note that here we use LLM as a judge to validate AVR-generated patches.
The PoC and Equivalence columns report the number of successful repairs on the PoC Validation and Equivalence Validation datasets, respectively.
To compare LLM-as-a-Judge with dynamic patch validation, we further compute their consistency rate on the PoC Validation dataset.
Across the \update{two LLM judges (GPT4.1 and Gemini2.5), the consistency rate ranges from 76.32\% to 81.14\%}. 
This level of consistency is expected, as LLM judges do not execute code and may therefore overlook compilation or runtime errors, such as the syntax error illustrated in Figure~\ref{llm_judge_case}.
Moreover, LLM judges occasionally produce false negatives, typically when an AVR-generated patch diverges syntactically from the ground-truth patch but still fulfills the same security objective.
In such cases, dynamic testing correctly identifies the patches as successful, while LLM-as-a-Judge may misclassify them as unsuccessful.

These findings suggest that dynamic patch verification remains the most reliable approach for evaluating repair correctness, as it executes PoCs and unit tests to confirm whether a vulnerability is truly fixed. 
However, dynamic verification is costly in terms of environment setup and test preparation. 
By contrast, LLM-as-a-Judge offers a lightweight alternative, achieving around 80\% consistency with dynamic testing. 
Thus, while dynamic patch verification should be considered the ground-truth standard, LLM-based judgment can serve as a practical and cost-effective substitute in settings where approximate yet efficient evaluation is sufficient.

\begin{figure}[t!]
    \centering
    \begin{subfigure}[b]{0.48\textwidth}
        \includegraphics[width=\textwidth]{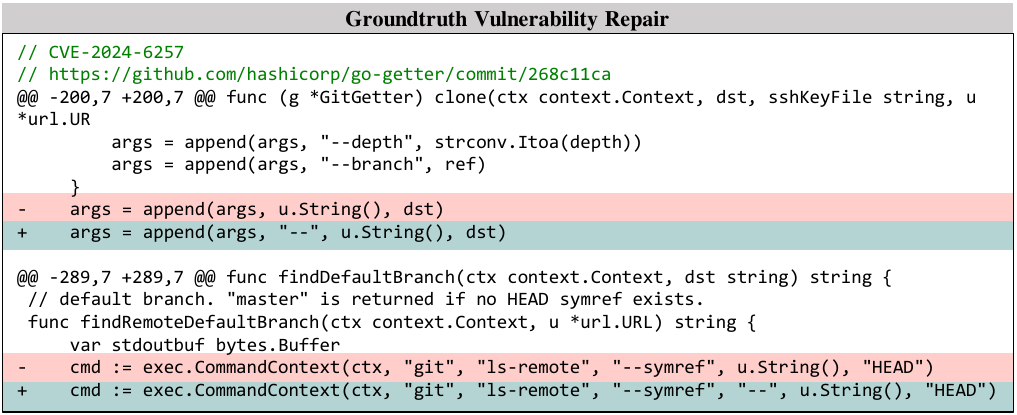}
        \caption{Groundtruth Patch for CVE-2024-6257}
        \label{new_case3_1}
    \end{subfigure}

    \begin{subfigure}[b]{0.48\textwidth}
        \includegraphics[width=\textwidth]{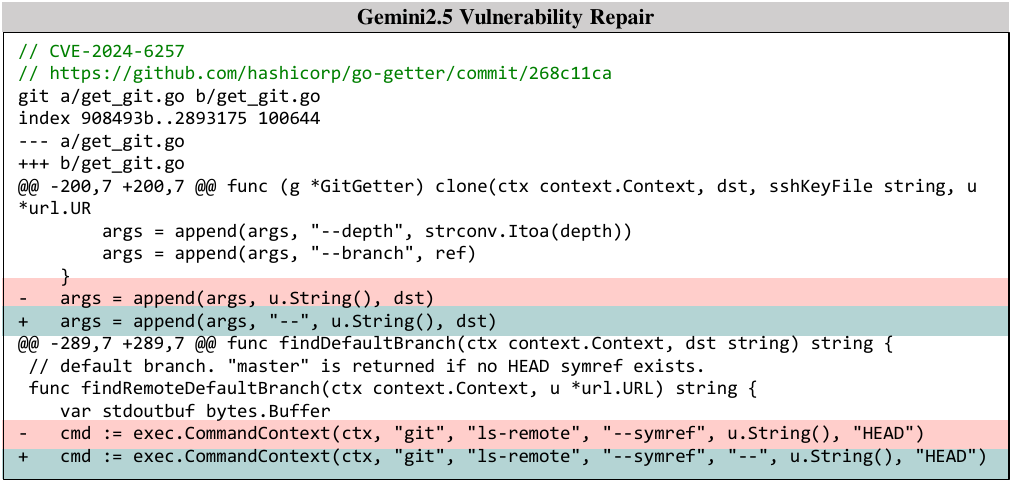}
        \caption{Successful Patch for CVE-2024-6257 Generated by Gemini2.5}
        \label{new_case3_2}
    \end{subfigure}

    \begin{subfigure}[b]{0.48\textwidth}
        \includegraphics[width=\textwidth]{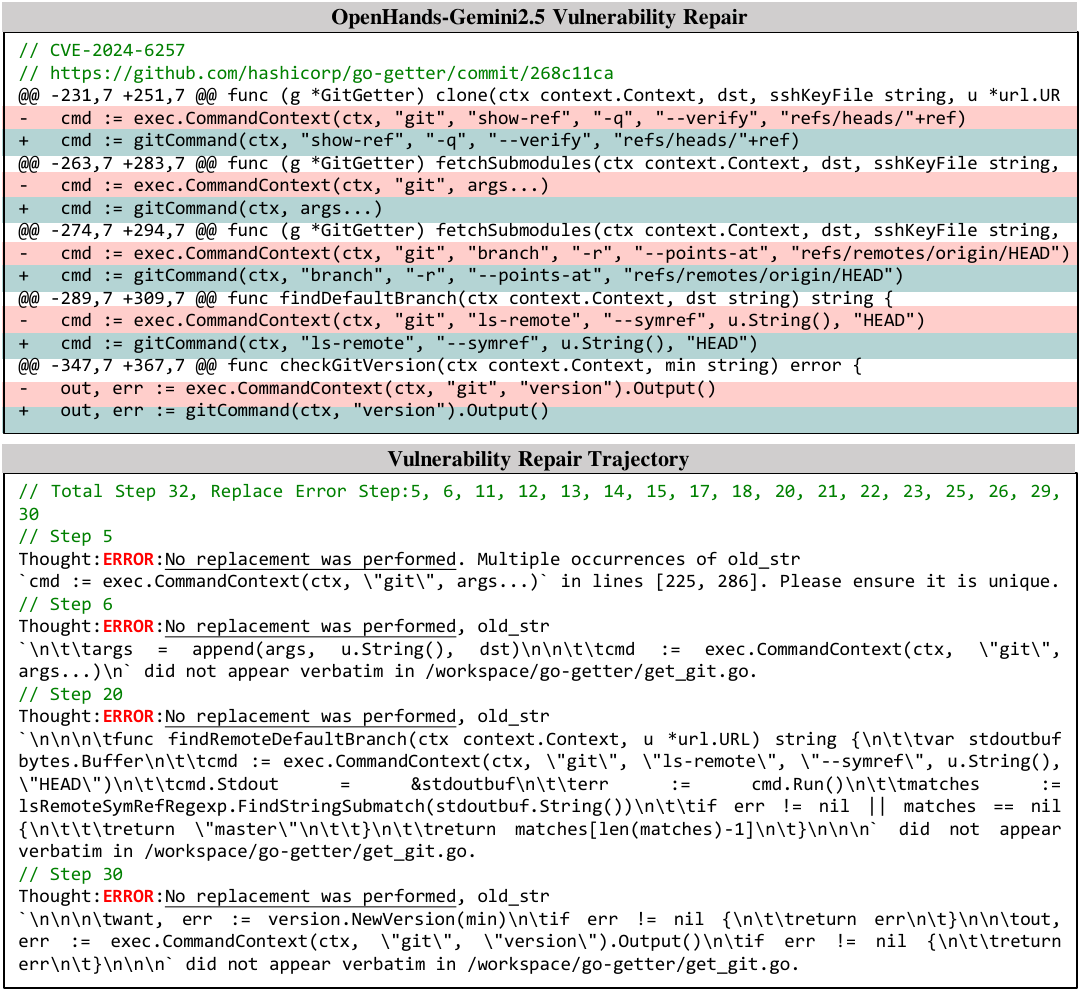}
        \caption{Error Steps while Generating the incorrect Patch for CVE-2024-6257 by OpenHands-Gemini2.5}
        \label{new_case3_3}
    \end{subfigure}
    \caption{\update{Patch Generation for CVE-2024-6257}}
    
    \label{new_case3}
\end{figure}
\begin{figure}[t!]
    \centering
    \begin{subfigure}[b]{0.48\textwidth}
        \includegraphics[width=\textwidth]{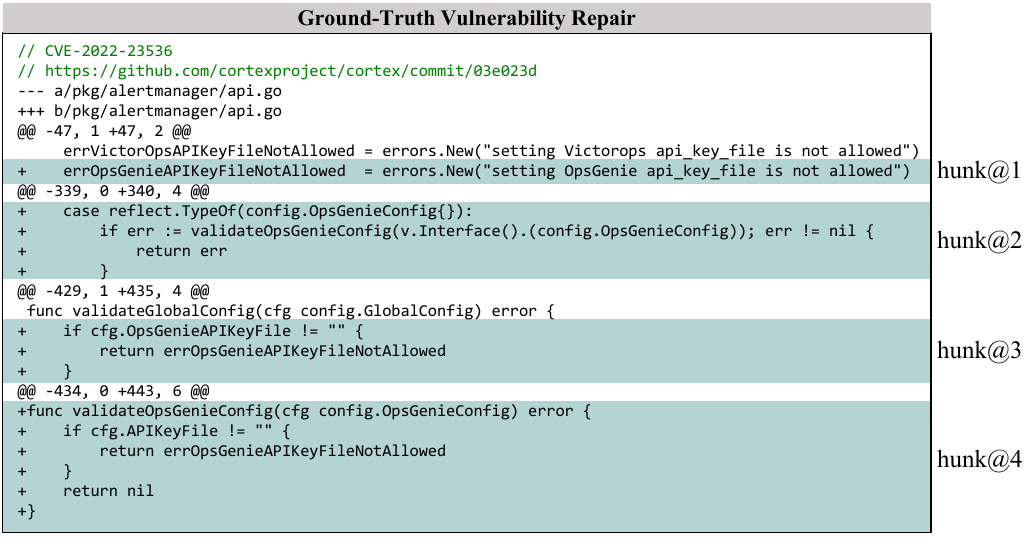}
        \caption{Ground-Truth Patch for CVE-2022-23536}
        \label{case4_1}
    \end{subfigure}

    \begin{subfigure}[b]{0.48\textwidth}
        \includegraphics[width=\textwidth]{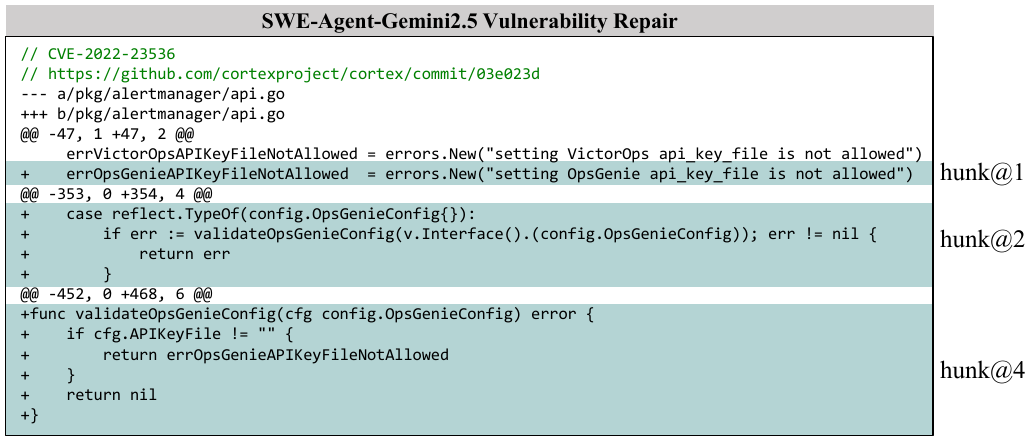}
        \caption{Failed Patch for CVE-2022-23536 Generated by \update{SWE-Agent-Gemini2.5}}
        \label{case4_2}
    \end{subfigure}
    \caption{Patch Generation for CVE-2022-22536}
    \label{case4}
\end{figure}

\subsection{Case Study: Repair CVE-2024-6257}
\label{case_study_agent_LLM}
\update{
Figure~\ref{new_case3} presents a CVE which the standalone LLM successfully repairs, whereas its associated agent fails.
Specifically, Gemini2.5 generates a patch that precisely matches the ground-truth fix.
In contrast, the patch generated by OpenHands-Gemini2.5 is not only incorrect but also considerably more divergent. 
This divergence mainly results from the agent’s full codebase visibility: when granted unrestricted navigation across the entire project, the agent frequently modifies code regions unrelated to the root cause of the vulnerability, resulting in unnecessary and counterproductive edits that ultimately degrade repair quality.
Additionally, OpenHands introduces a substantial number of string-replacement errors.
Across the 32 steps in its execution trajectory, 17 steps contain such mistakes, meaning that more than half of the agent’s actions are error-prone, which significantly undermines its ability to synthesize a valid patch.
}

\subsection{Case Study: Repair CVE-2022-22536}
\label{case_study_agent_LLM_2}
Figure~\ref{case4} presents a case study in which all approaches fail to generate a valid patch.
The ground-truth patch consists of four hunks that strengthen security checks for \texttt{api\_key\_file} in both the global and \texttt{OpsGenie} configurations.
However, all LLMs and agents fail because they omit the global validation in \texttt{hunk@3}.
This outcome is unsurprising, as generating multi-hunk patches is inherently challenging.
Such repairs require reasoning across a broader codebase and its dependencies, whereas current models tend to focus primarily on local context.

\subsection{Case Study: Repair CVE-2022-46146}
\label{case_study_agent_LLM_3}
\begin{figure}[t!]
    \centering
    \begin{subfigure}[b]{0.48\textwidth}
        \includegraphics[width=\textwidth]{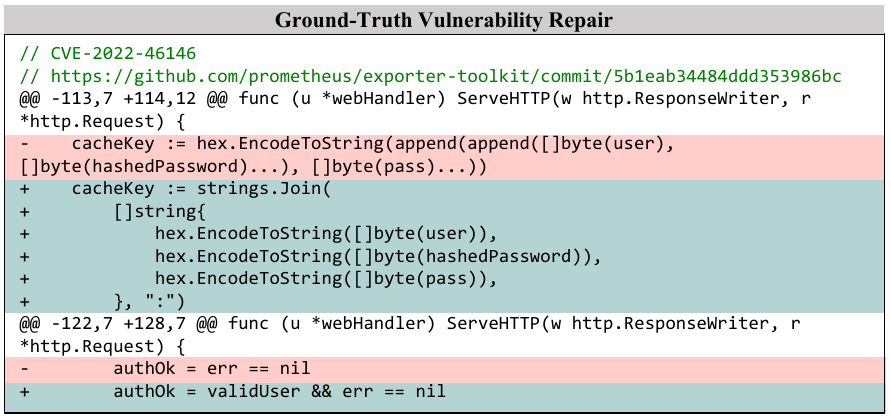}
        \caption{Ground-Truth Patch for CVE-2022-46146}
        \label{case5_1}
    \end{subfigure}

    \begin{subfigure}[b]{0.48\textwidth}
        \includegraphics[width=\textwidth]{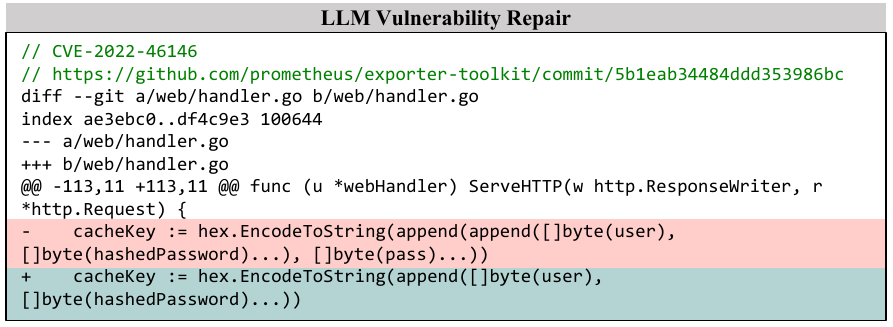}
        \caption{LLM Generated Patch for CVE-2022-46146}
        \label{case5_2}
    \end{subfigure}

    \begin{subfigure}[b]{0.48\textwidth}
        \includegraphics[width=\textwidth]{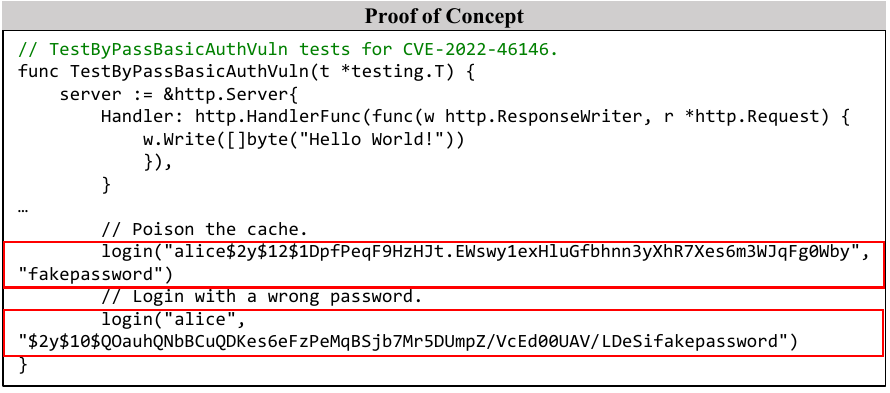}
        \caption{PoC for CVE-2022-46146}
        \label{case5_3}
    \end{subfigure}

    \begin{subfigure}[b]{0.48\textwidth}
        \includegraphics[width=\textwidth]{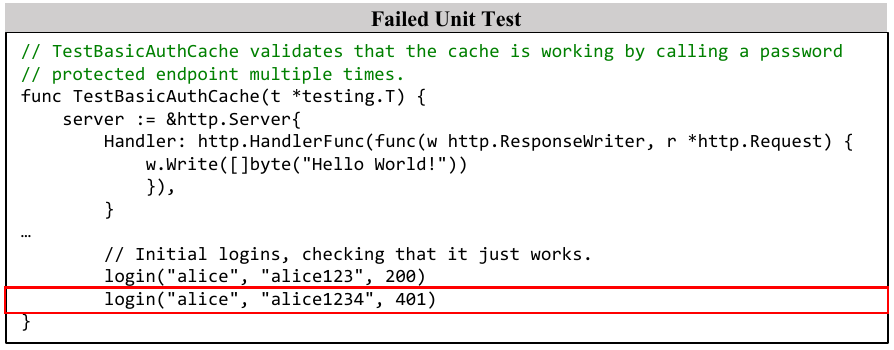}
        \caption{Failed Unit Test for CVE-2022-46146}
        \label{case5_4}
    \end{subfigure}
    
    \caption{Patch Generation for CVE-2022-46146}
    \label{case5}
\end{figure}
\update{
Figure~\ref{case5} presents a case study of CVE-2022-46146, a cache-poisoning authentication-bypass vulnerability.
As shown in Figure~\ref{case5_1}, the ground-truth fix introduces a delimiter (:) in cache-key construction and enforces a permission check, preventing forged credentials.
The PoC exploit (Figure~\ref{case5_3}) triggers the flaw via a crafted username and password.
Although the LLM-generated patch (Figure~\ref{case5_2}) passes the PoC, it removes the password component from the cache key, disrupting normal authentication and failing unit tests (Figure~\ref{case5_4}).
This case highlights that PoC-only patch validation can yield false-positive patches; effective validation should combine security tests and functionality tests to ensure that vulnerabilities are neutralized without introducing regressions.
}

\end{document}